\documentclass[useAMS,usenatbib]{mn2e}
\usepackage{times}
\usepackage{graphicx}
\newcommand{\mpcoh}{\,h^{-1}\,{\rm Mpc}}
\newcommand{\beq}{\begin{equation}}
\newcommand{\eeq}{\end{equation}}

\title[RSD with photometric data]
{Measuring Redshift-Space Distortions using Photometric Surveys}

\author[A. J. Ross et al.]{
  Ashley J. Ross\thanks{Email: Ashley.Ross@port.ac.uk}$^{1}$, 
  Will J. Percival$^{1}$, 
  Mart{\'i}n Crocce$^{2}$,
  Anna Cabr{\'e}$^{3}$, \& 
  Enrique Gazta{\~n}aga$^{2}$ \\
  $^1$Institute of Cosmology \& Gravitation, Dennis Sciama Building, 
    University of Portsmouth, Portsmouth, PO1 3FX, UK\\
  $^2$Institut de Ci\`encies de l'Espai, CSIC/IEEC, 
F. de Ci\`encies, Torre C5 par-2,  Barcelona 08193, Spain\\
  $^3$Center for Particle Cosmology, University of Pennsylvania, 
    209, South 33$^{rd}$ Street, Philadelphia, PA, 19104, USA\\}
\begin{document}

\date{Accepted by MNRAS}

\pagerange{\pageref{firstpage}--\pageref{lastpage}} \pubyear{2010}

\maketitle

\label{firstpage}

\begin{abstract}
  We outline how redshift-space distortions (RSD) can be measured from
  the angular correlation function $w(\theta)$, of galaxies selected
  from photometric surveys. The natural degeneracy between RSD and
  galaxy bias can be minimized by comparing results from bins with
  top-hat galaxy selection in redshift, and bins based on the radial
  position of galaxy pair centres. This comparison can also be used to
  test the accuracy of the photometric
  redshifts. The presence of RSD will be clearly detectable with the
  next generation of photometric redshift surveys. We show that the
  Dark Energy Survey (DES) will be able to measure $f(z)\sigma_8(z)$
  to a 1$\sigma$ accuracy of $(17\times b)$\%, using galaxies drawn
  from a single narrow redshift slice centered at $z=1$. Here $b$ is
  the linear bias, and $f$ is the logarithmic rate of change of the
  linear growth rate with respect to the scale factor. Extending to
  measurements of $w(\theta)$ for a series of bins of width
  $0.02(1+z)$ over $0.5<z<1.4$ will measure $\gamma$ to a 1$\sigma$
  accuracy of $25$\%, given the model $f = \Omega_m(z)^{\gamma}$, and
  assuming a linear bias model that evolves such that $b = 0.5 +
  z$ (and fixing other cosmological parameters). The accuracy of our analytic predictions is confirmed using mock
  catalogs drawn from simulations conducted by the MICE
  collaboration.
\end{abstract}


\section{Introduction}

The statistical analysis of redshift-space distortions (RSD) provides
constraints on the build-up of cosmological structure (see
\citealt{hamilton98} for a review). RSD are apparent anisotropic
patterns of galaxies formed when redshifts are converted into galaxy
distances assuming the Hubble flow: coherent comoving peculiar
velocities are translated into apparent distributions of galaxies. On
small distance scales ($\sim 1\mpcoh$), we see the well-known "finger
of god" effect (see, e.g., \citealt{Jackson72}), due to orbital
velocities of galaxies that are members of bound structures.  On
larger distance scales, galaxies move with structure growth, leading
to an apparent excess in the clustering strength along the
line-of-sight.

Following linear theory, the large-scale effect of RSD on 2-point statistics can be modelled as a
simple enhancement of the galaxy overdensity field
\citep{Kaiser}. This leads to an anisotropic correlation function
\citep{hamilton92} that is a function of both radial and transverse
separations. The excess line-of-sight clustering strength depends on
the amplitude of the matter velocity field that, following linear
theory, is usually parametrised by $f(z)\sigma_{8,{\rm mass}}(z)$,
where $f$ is the rate of change of the linear growth rate $f\equiv
d\log G/d\log a$, $G$ is the linear growth rate, $a$ is the scale
factor of the Universe, and $\sigma_{8,{\rm mass}}(z)$ is the root
mean square amplitude of fluctuations in the matter overdensity field
in spheres of radius $8\mpcoh$. We will now drop the ``mass'' label as
we will only refer to $\sigma_8(z)$ as normalising the matter
fluctuation field from here-on. Without the RSD, the amplitude of the
real-space galaxy overdensity field can be parametrised by
$b(z)\sigma_{8}(z)$, where $b(z)$ is the bias of the galaxies. For
$\Lambda$CDM cosmologies, to a good approximation, $f(z)
\equiv\Omega_m(z)^{\gamma}$ \citep{linder05} with $\gamma = 0.557$, and measurements of
$\xi^s({\bf r})$ can be used to test the standard cosmological model --- either by translating them into measurements of $f(z)$ using an external constraint on the galaxy bias or performing a dual fit for $b(z)\sigma_8(z)$ and $f(z)\sigma_8(z)$ (while accounting for the implied change in $\sigma_8(z)$ when $f(z)$ is changed).

RSD measurements have traditionally been confined to spectroscopic
redshift surveys, where accurate redshifts have been obtained for a
sample of galaxies: measurements from the 2dF Galaxy Redshift Survey
(2dFGRS; \citealt{colless03}) and the Sloan Digital Sky Survey (SDSS;
\citealt{york00}) are given by \citet{peacock01, hawkins03,
  percival04, pope04, zehavi05, okumura08, cabre09} for example, and
RSD have recently been detected at higher redshift
\citep{guzzo09,blake10}. 

 Photometric surveys are usually overlooked
when it comes to RSD measurements because the photometric redshift
errors are thought to wash out any RSD signal. Without question, these errors will significantly degrade the imprint of RSD. However, it is timely
to investigate the extent to which this is true given both the
high-quality of upcoming data from the Dark Energy Survey (DES {\tt
  www.darkenergysurvey.org}), the Panoramic Survey Telescope and Rapid
Response System (PanStarrs {\tt pan-starrs.ifa.hawaii.edu}), and the
Large Synoptic Survey Telescope (LSST {\tt www.lsst.org}), and recent
theoretical analyses that have investigated the effects of RSD on
projected clustering measurements.

If galaxy redshifts are used to select samples, then we see
redshift-space effects in angular/projected clustering measurements, even
though the angular positions of galaxies are unchanged
(\citealt{fisher93,nock10}, hereafter N10). Corrections for these
effects were recently applied to angular power-spectra analyses of the SDSS by
\citet{blake07}, \citet{padmanabhan07}, and \citet{Thomas10}; the latter of whom have used the corrections to place constraints on $f$. 

Working in configuration space, these corrections have
been shown to agree with $w(\theta)$ and its covariance calculated
with mock photometric redshift catalogs (N10,
\citealt{Crocce10}, hereafter C10). This effect arises because RSD impart coherent
fluctuations in galaxy density along any redshift-dependent boundary
imposed on the galaxies. Correct treatment of the redshift-space
distortions is extremely important, as studies suggest that the most
precise (non-RSD) cosmological constraints are obtained from using
narrow redshift bins \citep{sanchez10}, which also give
the strongest RSD signal (N10, C10).

N10 have shown that constructing galaxy samples using {\it
  pair-centre} binning removes the effects of RSD. Such
a binning scheme does not consider the individual redshift of a
galaxy, but rather the average redshift of (or distance to) the pair
of galaxies. Thus, a pair-centre binning scheme with $0.45 < z < 0.55$
will include all of the galaxy pairs with average redshifts between
0.45 and 0.55. In standard `top-hat' binning, all of the galaxies with
redshifts between 0.45 and 0.55 would be included. The symmetry
imparted by the pair centre binning is sufficient to remove the effect
of RSD.

RSD and distortions caused by using the wrong geometrical model when
analysing data in three dimensions (the Alcock-Palczynski (AP) effect
\citealt{alcock79}) give rise to similar distortions in the measured
clustering \citep{ballinger96,simpson09}, although the strength of the
degeneracy depends on the cosmological assumptions used in the data
analysis \citep{samushia10}. Thus, in general, analyses of RSD using
3D data either have to be performed simultaneously with geometrical
measurements, or utilise (and rely on) more precise geometrical
constraints. The measured angular correlation function, which only depends on
redshifts and angles, does not depend on the cosmological model to be
fitted. Fitting models to the measured angular clustering is
therefore considerably simpler. 

In this work we investigate the precision to which one can measure
cosmological structure growth using galaxy samples with photometric
redshifts.  In Section~\ref{sec:model}, we present the theoretical
modelling we employ to estimate angular correlation functions of
galaxies, $w(\theta)$ and their error, for different galaxy binning
schemes.  In Section~\ref{sec:mocks}, we test our analytic error predictions against mock
catalogs produced by the MICE simulation and show the importance of
including non-linear effects when measuring $f(z)\sigma_{8}(z)$.  In Section~\ref{sec:measureRSD}, we determine the precision to which $f(z)\sigma_{8}(z)$ can be measured for a survey similar to
DES using redshift bins of different sizes and mean redshifts and
galaxies with various mean photometric redshift errors and linear
bias.  In Section~\ref{sec:pc_bin}, we test the robustness of the measurements
against changes in the cosmology and systematics in the photometric
redshifts and we show how comparing measurements using standard binning schemes and those using pair-centre binning measurements can test the accuracy of the photometric redshifts. We conclude in Section~\ref{sec:conc}. Where
appropriate we assume a flat $\Lambda$CDM model with $\Omega_m =
0.25$, $h=0.7$, and $\Omega_b = 0.045$.

\section{Modeling Angular Correlation Functions} \label{sec:model}

For a sample of galaxy pairs, the angular-correlation function
$w(\theta)$, can be modeled by integrating the redshift-space
correlation-function $\xi^s({\bf r})$, over the joint distribution of
galaxy pairs $f(z_1,z_2)$, \citep{phil78}:
\begin{equation}  \label{eq:w2}
  w(\theta) = \int dz_1 \int dz_2 f(z_1,z_2)
    \xi^s\left[{\bf r}(\theta,z_1,z_2)\right],
\end{equation}
where the galaxy separation ${\bf r}$ is a function of the angular
separation of the galaxies $\theta$ and their redshifts $z_1$ and
$z_2$. For a binning scheme based on galaxy position (such as top-hat
binning), we can separate the selection function for the two galaxies. We can
then replace $f(z_1,z_2)$ by $n(z_1)n(z_2)$, where $n(z)$ is the normalized true redshift distribution, in the above expression. We note that, in practice, $n(z)$ is an estimate usually obtained from the photometric redshifts and their errors. We investigate potential systematics due to this estimation in Section \ref{sec:bdndz}.

For the pair-centre binning, $f(z_1,z_2)$ can be replaced by $n_{pc}(z_1)n_d(|z_1-z_2|)$, where $n_{pc}(z)$ is the normalized distribution of pair centres and $n_d(\Delta z)$ is the normalized distribution of pair separations (in redshift; note that the tophat $w(\theta)$ is recovered when integrating over $n_{pc}(z_1)n_d(|z_1-z_2|)$ when $z_1$ and $z_2$ are restricted to lie within the tophat bounds).  We incorporate the photometric redshifts into $f(z_1,z_2)$,
working with a redshift-space correlation function, rather than
photometric redshift-space one, so the functions $n(z)$ and
$f(z_1,z_2)$ depend on the photometric redshift error. For simplicity
these errors are assumed to be Gaussian with standard deviation
$\sigma_z$. In practice, this means that we Monte Carlo sample the true redshift distribution twice, and
then Monte Carlo sample the respective Gaussians to find two positions in photometric redshift space. If the pair-centre of these positions lies within the photometric redshift bin, the true redshift pair-centre and true redshift separation is added to its respective distribution (and the respective distributions are then normalized). See N10 for further details.

In this work, we only consider the plane-parallel approximation to the
geometry, and assume that $\xi^s(r)$ is given by \citep{hamilton92}
\begin{equation}
\xi^s(\mu,r) = \xi_o(r)P_o(\mu)+\xi_2(r)P_2(\mu) +  \xi_4(r)P_4(\mu),
\label{eq:ximus}
\end{equation}
where
\begin{eqnarray}
  \xi_o(r) &=& (b^2+\frac{2}{3}bf+\frac{1}{5}f^2)\xi(r), \\
  \xi_2(r) &=& (\frac{4}{3}bf+\frac{4}{7}f^2)[\xi(r)-\xi'(r)], \\
  \xi_4(r) &=& \frac{8}{35}f^2[\xi(r)+\frac{5}{2}\xi'(r)-\frac{7}{2}\xi''(r)],
\label{eq:xirs}
\end{eqnarray}
$P_\ell$ are the standard Legendre polynomials, and
\begin{eqnarray}
  \xi'\equiv3r^{-3}\int^r_0\xi(r')(r')^2dr' \\ 
  \xi''\equiv5r^{-5}\int^r_0\xi(r')(r')^4dr',
\end{eqnarray}
$b$ is the large-scale bias of the galaxy population being considered
and $\xi(r)$ is the isotropic 3-dimensional, real-space correlation
function.  The factor $\mu$ is the cosine of the angle between the
separation along the line of sight and the transverse separation. We
calculate $\xi(r)$ by Fourier transform from a power spectrum
calculated as described in \citet{EH98} and include the non-linear
modelling given in equation~10 of C10.  

The model we employ makes some strong assumptions (linear RSD, linear bias, weak non-linear gravity).  We note that its application to projected/angular correlation functions is well tested with simulation
(N10, C10) and has been shown to be a good fit to measured $w(\theta)$ (see, e.g., \citealt{R07,R08,Rred} for discussions of the range of scales over which measurements of angular clustering are well-fit by a linear bias model and, e.g., \citealt{sam11} for a discussion of the limits to which the linear RSD model is valid).

Each moment of $\xi(r)$ entering Eq.~\ref{eq:ximus} is normalised to
the standard correlation function, and hence its amplitude is fixed by
$\sigma^2_{8}(z)$. Any measurements of $b(z)$ or $f(z)$ calculated
using this model are therefore perfectly degenerate with the value of
$\sigma_{8}(z)$.  It is therefore most convenient for our purposes to
parametrize this model with two, separate, factors: $b(z)\sigma_{8}(z)$
and $f(z)\sigma_{8}(z)$ (e.g. \citealt{SP09}).

Eq.~\ref{eq:w2} shows that binning schemes act as ``moments'' of the
anisotropic redshift-space correlation function: each provides an
integral of this function weighted by a different distribution of
angles to the line-of-sight. The binning schemes therefore work in a
similar way to the Legendre polynomials in the standard definition of
the anisotropic correlation function. Comparing results from different
schemes such as considering measurements from both the top-hat and
pair-centre schemes, would allow the simultaneous measurement of both
$b(z)\sigma_{8}(z)$ and $f(z)\sigma_{8}(z)$ (when accounting for the covariance
between these measurements).

\subsection{Errors and Covariance}  \label{sec:cov}

Recent work has considered the problem of analytically calculating
covariances for angular/projected clustering measurements
(\citealt{cohn09}, C10). In order to set constraints on how well
future surveys will measure the angular clustering, we follow the
approach of C10, which is based on the angular power
spectrum for redshift slice $i$, which we denote $C_{\ell,i}$. 

The error on the full-sky angular power spectrum, $C_\ell$, is given by (see, e.g., \citealt{Dod03}) 
\begin{eqnarray}
\sigma^2(C_{\ell}) & \equiv & \langle C^2_\ell \rangle - \langle C_\ell \rangle\langle C_\ell \rangle \\
& = &\frac{2}{(2\ell+1)}\left(C_{\ell}+1/\bar{n}\right)^2
\label{eq:sigcl}
\end{eqnarray}
where $\bar{n}$ is number density of objects per steradian.
The predicted variance of the cross-power spectrum, $C_{\ell}^{i,j}$, between samples $i$ and $j$ is given by (see, e.g., eqs. 18 and 19 of \citealt{fish})
\begin{equation}
\begin{array}{ll}
\sigma^2(C_{\ell}^{i,j}) =& \frac{1}{(2\ell+1)}\\
 & \times \left[(C_{\ell}^{i,j})^2+(C_{\ell}^{i}+1/\bar{n}_i)(C_{\ell}^{j}+1/\bar{n}_j)\right].
\label{eq:sig2cl}
\end{array}
\end{equation}
Thus, given that 
\begin{equation}
  w_{i,j} (\theta) = \sum_{\ell \geq 0} \left(\frac{2\ell+1}{4\pi}\right)
    P_{\ell}(cos\theta)C_{\ell}^{i,j},
   \label{eq:clw}
\end{equation}
and that the effect of only observing a fraction of the sky, $f_{sky}$, is well approximated by simply dividing $\sigma^2(C_{\ell})$ by $f_{sky}$ (see, e.g., C10), the covariance (that we denote {\rm Cov}) between measurements of the angular cross-correlation function at $\theta$ and
$\theta^{\prime}$ between redshift slices $i$ and $j$ (that is the auto-correlation in the case $i=j$) can be modelled as
%\begin{equation}
%\begin{array}{lll}  \label{eq:clcov}
%  V_{\theta,i,\theta^{\prime},j} &\equiv& 
%    \langle w_i(\theta)w_j(\theta')\rangle
%    -\langle w_i(\theta)\rangle\langle w_j(\theta')\rangle\\
% & =&\frac{1}{f_{sky}}\sum_{\ell \geq 0} 
%    \frac{2\ell + 1}{(4\pi)^2}P_{\ell}({\rm cos}\theta)
%    P_{\ell}({\rm cos}\theta^{\prime})\\
%& & \times   \left[(C^2_{\ell,i,j}+(C_{\ell,i}+1/\bar{n}_i)(C_{\ell,j}+M/\bar{n}_j)\right]\nonumber\\ 
%&   & \hspace{1cm}+ 1/n_p\delta_{\theta,\theta^{\prime}},
\begin{eqnarray} 
&  {\rm Cov}_{i,j,\theta,\theta^{\prime}} \equiv
    \langle w_{i,j}(\theta)w_{i,j}(\theta')\rangle
    -\langle w_{i,j}(\theta)\rangle\langle w_{i,j}(\theta')\rangle  \label{eq:cov}\\
 & =\sum_{\ell \geq 0} 
    \frac{(2\ell + 1)^2}{f_{sky}(4\pi)^2}P_{\ell}({\rm cos}\theta)
    P_{\ell}({\rm cos}\theta^{\prime}) \sigma^2(C_{\ell}^{i,j})+ \frac{\delta_{\theta,\theta^{\prime}}}{n_p}  \label{eq:clcov}
\end{eqnarray}
%\end{equation}
where $n_p$ is
the number of galaxy pairs (that can be determined from the number
density and angular size of the bin) and $\delta_{\theta,\theta^{\prime}}$ is the Kronecker delta. 

Given $P(k)$, the $C_{\ell,i,j}$ spectra can be calculated via (see, e.g., \citealt{fisher93,padmanabhan07})
\begin{equation}
  C_{\ell}^{i,j}  = \frac{2}{\pi}\int k^2 dk P(k) \Psi_{\ell,i}(k)\Psi_{\ell,j}(k),
\end{equation}
where $\Psi_{\ell,i}$ is given by
\begin{equation}
  \Psi_{\ell,i}(k) = \Psi^r_{\ell,i} + \int dz n_i(z) D(z) j_{\ell}(k\chi(z)),
\end{equation}
and $j_{\ell}$ is the spherical Bessel function of order $\ell$ and
\begin{equation}
  \begin{array}{l}
  \Psi^r_{\ell,i}(k) = f/b\int dz\,n_i(z) D(z)\left[\frac{(2\ell^2+2\ell-1)}
    {(2\ell+3)(2\ell-1)}j_{\ell}(k\chi(z)) \right. ~\\ 
  - \left. \frac{\ell(\ell-1)}{(2\ell-1)(2\ell+1)}j_{\ell-2}(k\chi(z)) 
  - \frac{(\ell+1)(\ell+2)}{(2\ell+1)(2\ell+3)}j_{\ell+2}(k\chi(z))\right].
\end{array}
\end{equation}
This set of equations provides the full covariance matrix in both
scale and redshift, as required in order to make measurements of
$f(z)\sigma_{8}(z)$ at multiple redshifts.

\subsection{Pair-Centre Covariance}
\label{sec:pcerr}
For the pair-centre binning, determining the covariance is not as
straightforward. We can produce an approximate solution by calculating
the variances of all of the thin slices in photometric redshift that would contribute to the
pair-centre measurement and all of the co-variances between these slices. 
Given a photometric redshift survey, one can split it into very narrow consecutive shells
of width $\delta s$ in photometric redshift located at 
\beq
s_i = \delta s \times i   \ \ \ (i=0,\infty).
\eeq
If one wishes to estimate the pair-centre angular correlation function, $w_{pc}$, in a bin
centered at $s_b$ of width
$\Delta s_b$ (in photometric redshift space), the condition for a pair of galaxies lying in bins $i$
and $j$ to enter the selection is
\beq
y_{ij} = |\bar{s}_{ij}-s_b| / \Delta s_b \le 1
\eeq
with $\bar{s}_{ij}=\delta s \times (i+j)/2$ the mean photo-z of the galaxy
pair. Thus the condition can be implemented by a $\Theta(y)$ function
giving 1 if $y\le 1$ and 0 otherwise.
\newline

The pair-centre correlation can be estimated as,
\beq
w_{pc}(\theta) = \frac{ \sum_i \sum_{j \geq i} n(z_i) n(z_j) w_{i,j}(\theta) \Theta(y_{ij}) }{ \sum_i \sum_{j \geq i} n(z_i) n(z_j) \Theta(y_{ij})}
\label{eq:wpc}
\eeq
where $w_{i,j}(\theta)$ is the angular cross-correlation function between bins $i$, $j$ and $n(z_i)$ is the value of the overall redshift distribution $n(z)$ (i.e., the redshift distribution of sources over the entire survey being considered) at redshift $z_i$ with $C^{i,j}_{\ell}$ calculated as described by Eqs. 14-16, with $n_i(z)$ in Eq. 15 being the true redshift distribution of the thin slice in photometric redshift space, and the covariance of $w_{i,j}(\theta)$, ${\rm Cov}_{i,j}(\theta,\theta^{\prime})$, is therefore given by Eq. \ref{eq:clcov}. Thus, one can calculate all of the ${\rm Cov}_{i,j}(\theta,\theta^{\prime})$ contributing to the sum in Eq. \ref{eq:wpc}. In order to determine ${\rm Cov}_{pc}(\theta,\theta^{\prime})$, one must account for the covariance between $w_{h,i}(\theta)$ and $w_{j,k}(\theta)$, which we denote ${\rm Cov}_{h,i,j,k}$. We can estimate ${\rm Cov}_{h,i,j,k}$ by calculating the ``overlap", $O_{h,i,j,k}$, of the four thin photometric redshift shells,
\beq
O_{h,i,j,k} = \frac{\int dz n_h(z)n_i(z)n_j(z)n_k(z)}{\int dz n_h^2(z)n_i^2(z)} ,
\eeq
and assuming
\beq
{\rm Cov}_{h,i,j,k} = O_{h,i,j,k} {\rm Cov}_{h,i}(\bar{n} = \infty)
\eeq
where we set $\bar{n} = \infty$ since the covariance between the separate $C_{\ell}$ measurements should not have a shot-noise contribution. We thus estimate

\begin{equation}
\begin{array}{l}
 {\rm Cov}_{pc} = \frac{1}{W_t} \sum_{i,j \geq i} n^2(z_i) n^2(z_j) {\rm Cov_{i,j}} \Theta(y_{ij}) + \\
\frac{1}{W_t} \sum_{h, i \geq h,j,k > i} n(z_h) n(z_i) n(z_j)n(z_k){\rm Cov_{h,i,j,k}} \Theta(y_{hi}) \Theta(y_{jk})
\label{eq:vpc}
\end{array}
\end{equation}
where we use $\sum_{i,j}$ as shorthand for $\sum_i\sum_j$ and
\beq
W_t = \sum_{h, i \geq h,j,k > i} n(z_h) n(z_i) n(z_j)n(z_k) \Theta(y_{hi}) \Theta(y_{jk}).
\eeq
This approximation is validated by comparisons with mock catalogs in section \ref{sec:mocks}.

We estimate the covariance between the top hat and pair-centre binning by assuming the correlation, ${\rm Cov}_{pc,TH}/\sqrt{{\rm Cov}_{pc}{\rm Cov}_{TH}}$, between the measurements is equal to the ratio of the effective volume (given by Eq. 5 of \citealt{tegfish}) of the galaxies used in each binning scheme.  We find this correlation varies between $1/7$ and $1/2$, depending on the specific bin widths and median redshift. We find that these values are consistent with the covariance we find in the mocks used in Section \ref{sec:pccov}, though the mocks do exhibit some scale dependence. Allowing the covariance to
change by as much 25\% does not affect any of the conclusions of this work, but this issue is worthy of
further study in the future.

\begin{figure}
\includegraphics[width=0.9\columnwidth]{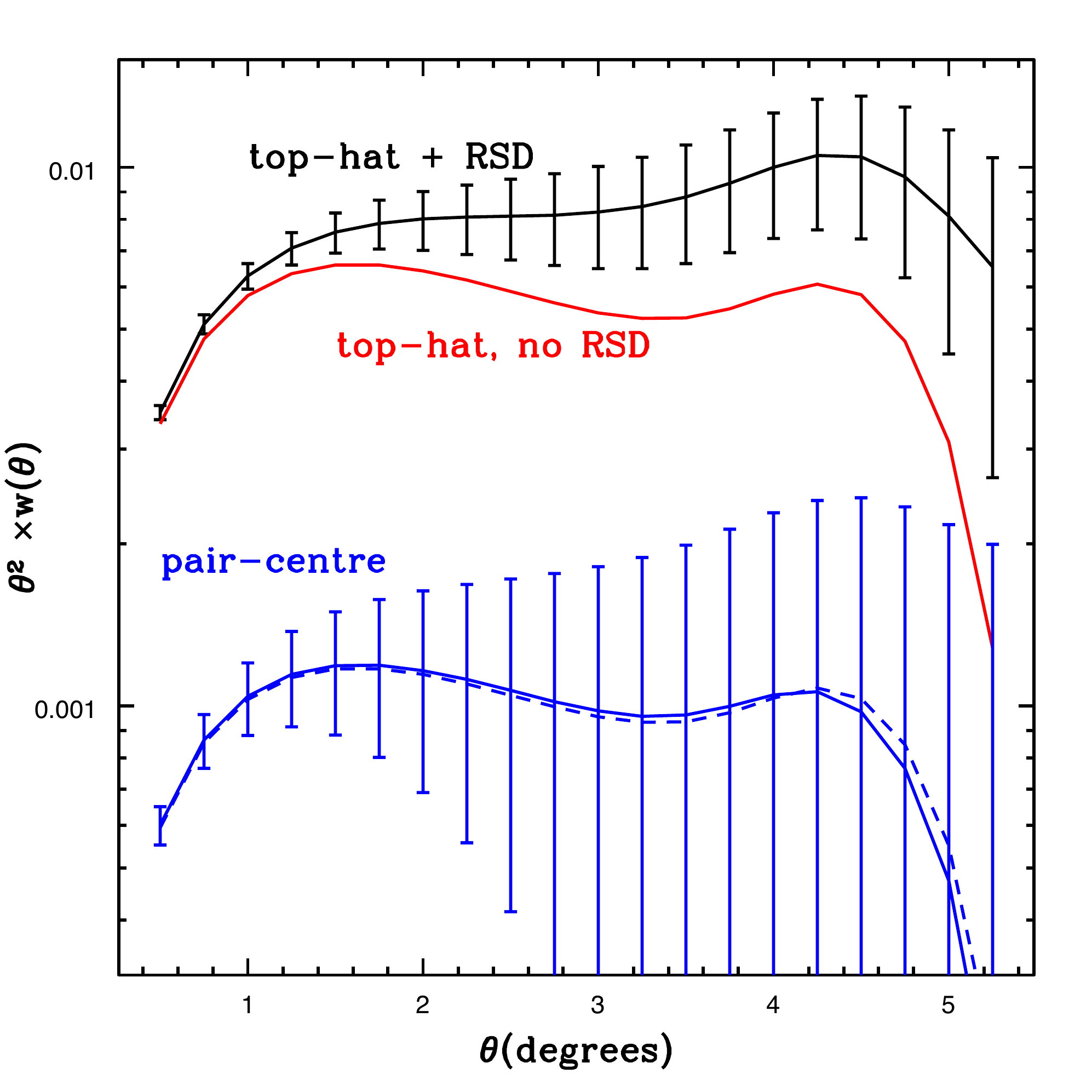}
\caption{Model angular correlations functions (multiplied by
  $\theta^2$ for clarity), calculated directly using Eq.~(\ref{eq:w2})
  from the 3D correlation function. These are for distributions of
  galaxies within the photometric redshift bin $0.45 < z < 0.55$ and
  assume $\sigma_z = 0.03(1+z)$. The black line is for a top-hat bin
  when redshift-distortions (RSD) are included, its and theoretical uncertainties are also displayed (calculated  via Eq. \ref{eq:clcov})
  , while the red line was calculated without RSD. The solid blue line is for a
  pair-centre binning, with its theoretical uncertainties (calculated via Eq. \ref{eq:vpc}) also displayed. The dashed-blue line displays the top-hat, no RSD prediction divided by 5.62. }
\label{fig:omcl}
\end{figure}

\subsection{Estimating Errors for DES}

As a specific example, we now consider the sensitivity of the angular
correlation functions expected to be measured for DES. As in N10, we
take the overall selection of DES galaxies to have the form
\begin{equation}  \label{eq:nzdes}
  n_{DES}(z) \propto (\frac{z}{0.5})^2exp(-\frac{z}{0.5})^{1.5},
\end{equation} 
and assume that DES will observe 300 million galaxies over 1/8th of
the sky\footnote{We thank the DES LSS working group for providing
  these estimations.}. These galaxies are assumed to be unbiased on
average, that is $b=1$. We will consider the dependence of our
conclusions on bias in Section~\ref{sec:gal_sample}.

Because the DES galaxy density is high, the predicted error for
clustering measurements at a given equivalent physical scale (i.e, $\sim \chi(z)\theta$, where $\chi(z)$ is the comoving distance to redshift $z$), for bins
that are of the same order as $\sigma_z$, will be driven almost
entirely by the median redshift of the sample: the shot noise term in
Eq.~\ref{eq:clcov} has a negligible impact on the final error. Thus,
for DES, ${\rm Cov}_{\theta,\theta^{\prime}} \propto C^2_{\ell}$, and we would
expect the same percentage error regardless of the width of the
redshift bin for both top-hat and pair-centre binning. Note that this
argument will break down around the BAO scale where the angular
correlation crosses from positive to negative and consequently the
proportionality arguments do not hold. Also, moving to smaller bins
will not necessarily provide significant additional signal because
results will be correlated through the photometric redshift errors.

We assume Gaussian photometric redshift errors, $\sigma_z$, of
$0.03(1+z)$. Although the average $\sigma_z$ of the full sample of DES
galaxies at these redshifts is predicted to be closer to $0.05(1+z)$ than $0.03(1+z)$ (see \citealt{ban08}, Fig. 1 of C10), a slight reduction
in the number of galaxies used can have a large effect on the mean
value of $\sigma_z$. As discussed above, the density of galaxies is
high so results are cosmic variance and photometric redshift limited: we can
therefore reduce the sample density without a significant increase in
noise. But the ability to measure RSD does depend strongly on the
width of the redshift distribution: Consequently, galaxy samples
should be constructed to have minimal average $\sigma_z$ (by
selecting, for example, primarily early-type galaxies). The ease of
redshift determination for these galaxies can lead to significant
gains: SDSS galaxies with $0.2 < z < 0.3$ and $M_r < -21.2$ have an
average $\sigma_z$ of 0.041, but if one removes all of the galaxies
with $\sigma_z$ greater than $0.053(1 + z)$, this is reduced to 0.03
and 82\% of the galaxies are left in the sample. (This sample was studied
by \citealt{R10}, and the estimated photometric redshift errors were consistent
with the measured clustering.)

Fig.~\ref{fig:omcl} compares model correlation functions (multiplied
by $\theta^2$ for clarity), calculated by directly projecting the 3D
correlation function. These were
calculated for galaxies within a photometric redshift bin $0.45 < z <
0.55$ and $\sigma_z = 0.03(1+z)$. We plot models for a top-hat bin in
real-space (red) and redshift-space (black), and for pair-centre binning (blue). We also display the theoretical uncertainties, as predicted
via Eq. \ref{eq:clcov} for the top-hat bin and Eq. \ref{eq:vpc} for the pair-centre
bin. The signal-to-noise is considerably worse for the pair-centre binning.

The dashed line in Fig.~\ref{fig:omcl} represents the case where we divide the real-space top-hat prediction by 5.62. This illustrates that the shape of the pair-centre measurement
is nearly identical to that of the top-hat without redshift space distortions. We
find this to be true in general when the width of the pair-centre
bin matches the width of the top-hat bin. While we find we can apply
slightly more narrow redshift bounds to the pair-centre bin and obtain
models that are slightly better matched, for simplicity, we will
always apply the same redshift bounds when comparing model pair-centre
and top-hat $w(\theta)$.

\begin{figure}
\includegraphics[width=84mm]{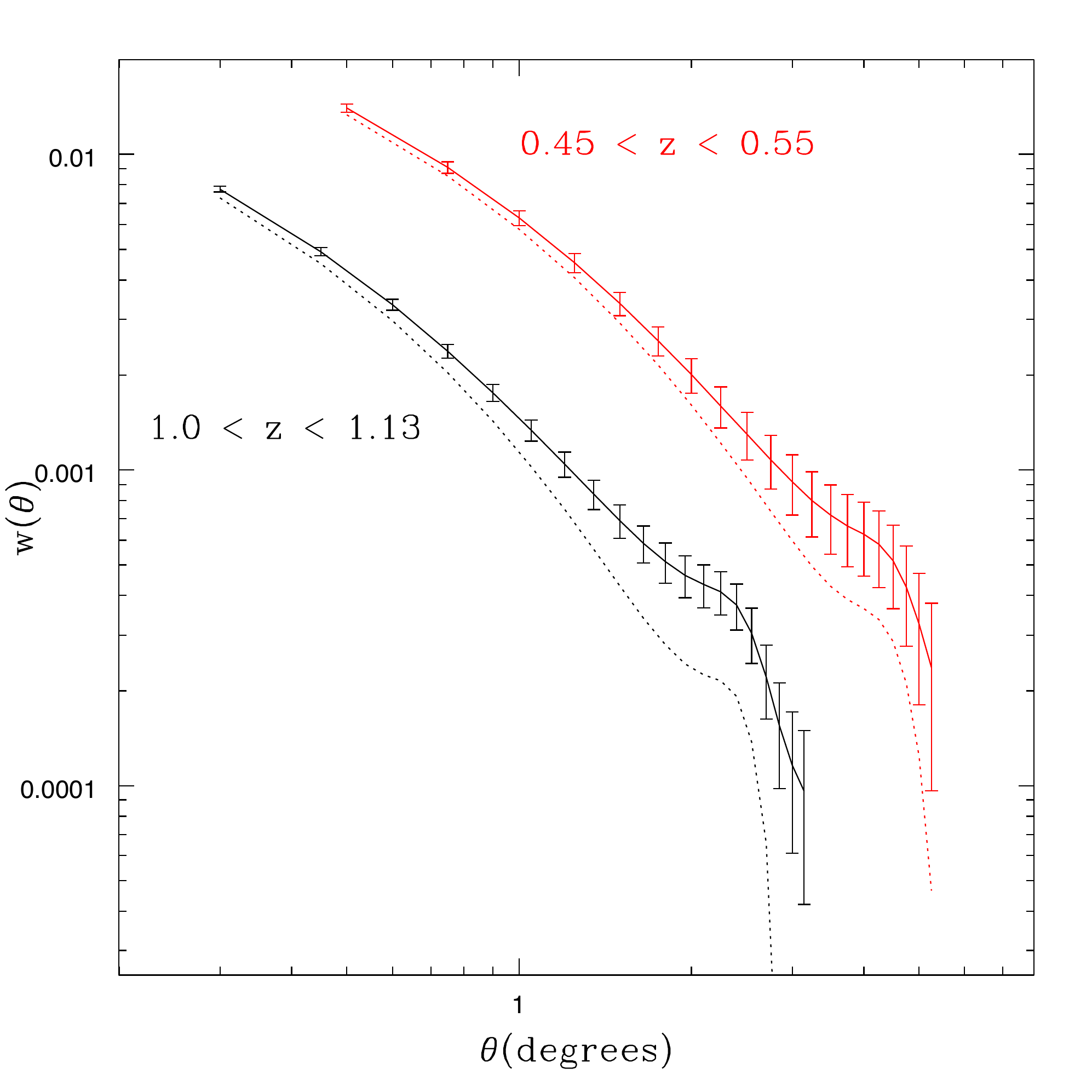}
\caption{Model angular correlation functions for galaxies drawn from
  the DES in top-hat bins with photometric redshifts $1.0<z_{\rm
    phot}<1.13$ (black) and $0.45<z_{\rm phot}<0.55$ (red).  The solid
  curves include RSD, dashed lines do not. Error bars, which are only
  plotted for models with RSD, were calculated using
  Eq.~(\ref{eq:clcov}). \label{fig:TH.5}}
\end{figure}
Fig.~\ref{fig:TH.5} presents the predicted angular correlation
functions for DES, with and without RSD, for top-hat photometric redshift bins centred at $z = 0.5$ and $z = 1.065$. For clarity, we only present
errors for the models with RSD.  Data are plotted for two redshift
bins, and we clearly see that the increase in volume surveyed at high
redshifts leads to more precise clustering measurements.

\section{Testing With Mocks}  \label{sec:mocks}

This section outlines tests of our analytic prediction for the
model $w(\theta)$ and its covariance matrices, performed using mock catalogs drawn from the
N-body simulation MICE7680 produced by the MICE
collaboration. This simulation
assumed a flat $\Lambda$CDM cosmology with parameters $\Omega_m =
0.25, \Omega_b = 0.044, \sigma_{8, {\rm mass}}(0) = 0.8, n_s = 0.95$ and
$h = 0.7$ and tracked the gravitational evolution of 2048$^3$
dark-matter particles within a comoving volume with a box size given
by $L_{{\rm box}} = 7680\mpcoh$ (see \citealt{Fosalba} and
\citealt{MICE} for further details).  We use mock catalogues corresponding to dark matter particles distributed in slices of varying width centered at the mean redshift of z=0.5  (i.e. extracted from the co-moving output at this z) and subtending 1/8 of the sky. The very large volume of MICE7680 allow us to obtain at least 125 independent mocks for every bin-width and photometric error under consideration\footnote{Mock catalogs are publicly available at http://www.ice.cat/mice}. We refer the reader to C10 for a detailed presentation of the mock build-up.

\subsection{Pair-centre Covariance}
\label{sec:pccov}
\begin{figure}
\includegraphics[width=84mm]{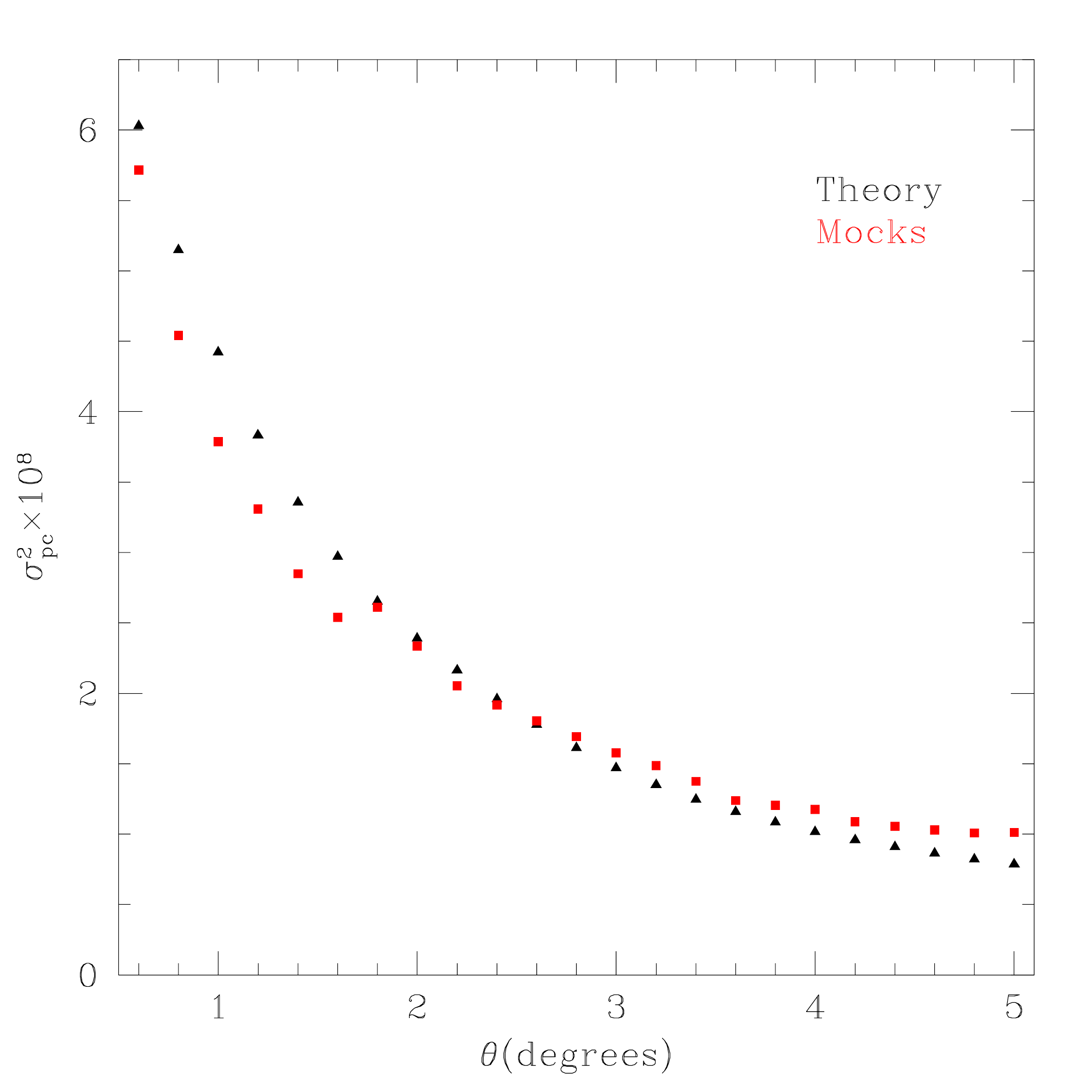}
\caption{The variance ($\sigma^2$) of the angular correlation function in the pair-centre bin $0.475 < z_{pc} < 0.525$ for objects with a DES-like distribution, determined from 125 mock realizations (black triangles) and estimated theoretically (using Eq. \ref{eq:vpc}) \label{fig:mockthpc}.}
\end{figure}

We use the mocks to test our estimation of the pair-centre covariance matrix. We note that the theoretical modelling of $w_{TH}$ and its covariance matrix has been extensively tested against the mocks in C10. Using a range in redshift
$0.2 < z < 0.8$, the volume was split, in photometric-redshift-space, into shells $15h^{-1}$Mpc thick. We then calculated $w_{pc}(\theta)$ for each realization by measuring all of the correlation- and cross-correlation functions of shells where $0.475 < (z_i+z_j)/2 < 0.525$, where $z_i$ is the mean redshift of slice $i$, and then summing them according to the overall redshift distribution. These measurements allow us to calculate the covariance matrix for $w_{pc}(\theta)$ for the pair-centre bin $0.475 < z_{pc} < 0.525$. Fig. \ref{fig:mockthpc} presents the diagonal elements of the covariance matrix, multiplied by $10^8$, determined from the mocks (red squares) and from Eq. \ref{eq:vpc} using shells 0.01 thick in photometric-redshift-space (black triangles). The theory appears to slightly underestimate the variance on large scales and slightly over-estimate the variance on small scales. However, the agreement is close enough that we should be able to use Eq. \ref{eq:vpc} for the purposes of this study. We find that the agreement between off-diagonal terms is quite similar, in terms of the absolute differences between the theoretical and mock estimates.

\subsection{Constraining $f(z)\sigma_8(z)$}
While the model $w_{TH}$ and its covariance matrix has already been shown to agree with the mocks (see C10), we wish to confirm that we can recover an unbiased value of $f(z)\sigma_8(z)$ and its uncertainty. We have tested four mock samples against the analytic predictions:
two that assume no photometric redshift errors, with widths of
$\Delta z = 0.05(1+z)$ and $\Delta z = 0.15(1+z)$ and two that have
redshift errors $0.02(1+z)$ and $\Delta z = 0.03(1+z)$ and $\Delta z = 0.05(1+z)$ (all have RSD effects applied). We only
test the predictions at medium redshift ($z\sim0.5$), where we can select
125 non-overlapping samples from the MICE simulations. While we are testing the results at $z = 0.5$, we expect the DES to obtain much more precise constraints at higher redshifts. However, given that the distribution of matter becomes more non-Gaussian as the Universe evolves, we can reasonably expect that the modelling that works at $z \sim 0.5$ should work at higher redshift as well. Thus, by confirming that our modelling is correct at $z \sim 0.5$, we expect our modelling to be correct in general. 

The mocks provide measurements of $w(\theta)$ with $0.2^{\rm o} \leq
\theta \leq 8^{\rm o}$ in angular bins of width $0.2^{\rm
  o}$. Jack-knife errors, $\sigma_{jk}(\theta)$, (see, e.g., \citealt{Scr02,Mye07,R07}) are
calculated for each of these 125 mocks, using 63 independent regions of equal area. In order to estimate the covariance matrix of each
realization, we determine the covariance matrix of the ensemble of mocks, ${\rm Cov}_{e}$, and use the covariance matrix, ${\rm Cov}_{jk}$ given by ${\rm Cov}_{jk}(\theta,\theta^{\prime}) = {\rm Cov}_{e}(\theta,\theta^{\prime})\sigma_{jk}(\theta)\sigma_{jk}((\theta^{\prime})/\sqrt{{\rm Cov}_{e}(\theta,\theta){\rm Cov}_{e}(\theta^{\prime},\theta^{\prime})}$. We find that, in the case of the most extreme fluctuations from the mean, the size of the jack-knife errors of individual angular bins are correlated with the size of the fluctuation. Modulating the covariance matrix based on the jack-knife errors thus reduces the weight of the more extreme fluctuations and we find this yields results matching our analytic predictions.
\begin{figure}
\includegraphics[width=84mm]{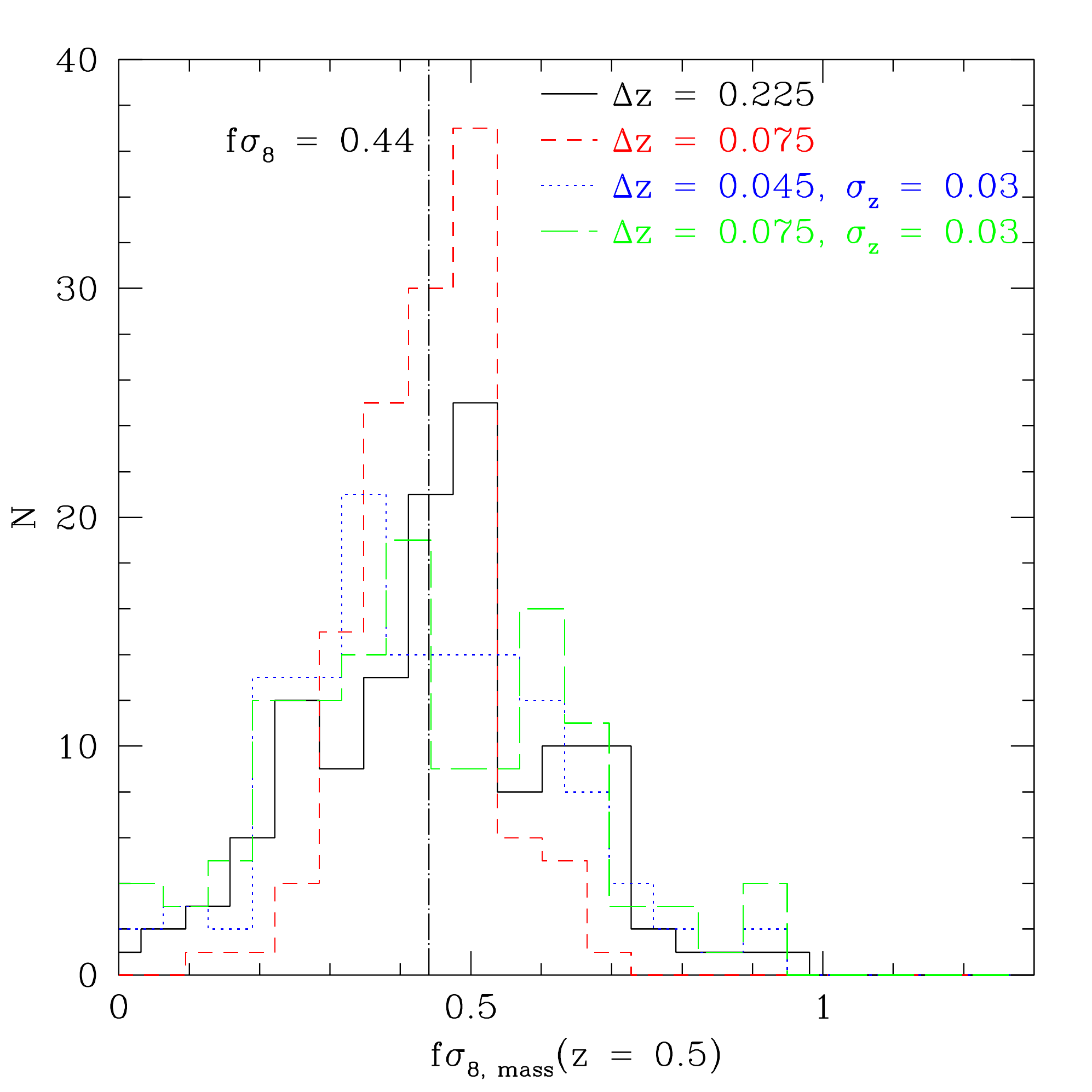}
\caption{Histograms displaying the distribution of $f(z)\sigma_{8}(z)$
  values obtained from 125 MICE mock catalogs with $\bar{z} = 0.5$ and
  $\Delta z = 0.15(1+z)$ (solid black line) and $\Delta z = 0.05(1+z)$ (dashed red line)
   with no redshift error, for a mock catalog with $\Delta z
  = 0.03(1+z)$ and redshift errors $0.02(1+z)$ (dotted blue line), and for a mock catalog with $\Delta z
  = 0.05(1+z)$ and redshift errors $0.02(1+z)$ (long-dashed green line). }
  \label{fig:micehist}
\end{figure}

To find the best-fit values of $f(z)\sigma_8(z)$, we fit for data in the range $1^{\rm o} \leq \theta \leq
5^{\rm o}$, {\bf marginalizing over $b(z)\sigma_8(z)$}. In order to produce the models we compare to these mocks, we use the same power-spectrum used by the MICE collaboration
(available at the MICE website). Fig.~\ref{fig:micehist} presents histograms of these best-fit
$f(z)\sigma_{8}(z)$ values obtained for four types of mock; $\Delta
z = 0.15(1+z)$ in solid black, $\Delta z = 0.05(1+z)$ in dashed red, $\Delta z
= 0.03(1+z)~ \& ~\sigma_z=0.02(1+z)$ in dotted blue, and $\Delta z
= 0.05(1+z) ~\& ~\sigma_z=0.02(1+z)$ in long dashed green.  For the mean and standard
deviation of the two distributions without redshift errors, we find 0.47$\pm$0.09 for $\Delta z =
0.05(1+z)$ and 0.48,$\pm$0.15 for $\Delta z =
0.15(1+z)$.  For $\sigma_z = 0.02(1+z)$, we find 0.42$\pm$0.18 for $\Delta z =
0.03(1+z)$ and 0.43$\pm$0.2 for $\Delta_z = 0.05(1+z)$. 

We make analytic predictions, based on the modelling described in Section \ref{sec:model}, using methods described fully in Section \ref{sec:measureRSD}. These predictions are $f(0.5)\sigma_{8}(0.5) = 0.44\pm0.09$,
$f(0.5)\sigma_8(0.5) = 0.44^{+0.13}_{-0.16}$, $f(0.5)\sigma_{8}(0.5)
= 0.44^{+0.16}_{-0.18}$, and $f(0.5)\sigma_{8}(0.5)
= 0.44^{+0.18}_{-0.20}$, respectively, for the four redshift bins
described above. These agree quite well with the results from the mocks and therefore suggest that our analytic formalism can be used to effectively predict the precision
to which $f(z)\sigma_{8}(z)$ can be measured with a DES-like survey. We further note that we have
applied a linear RSD model (given by Eqs. 2-7) to obtain all of these results, and this does not appear to bias the results.

\subsection{Importance of Non-linear Gravitational effects}

\begin{figure}
\includegraphics[width=84mm]{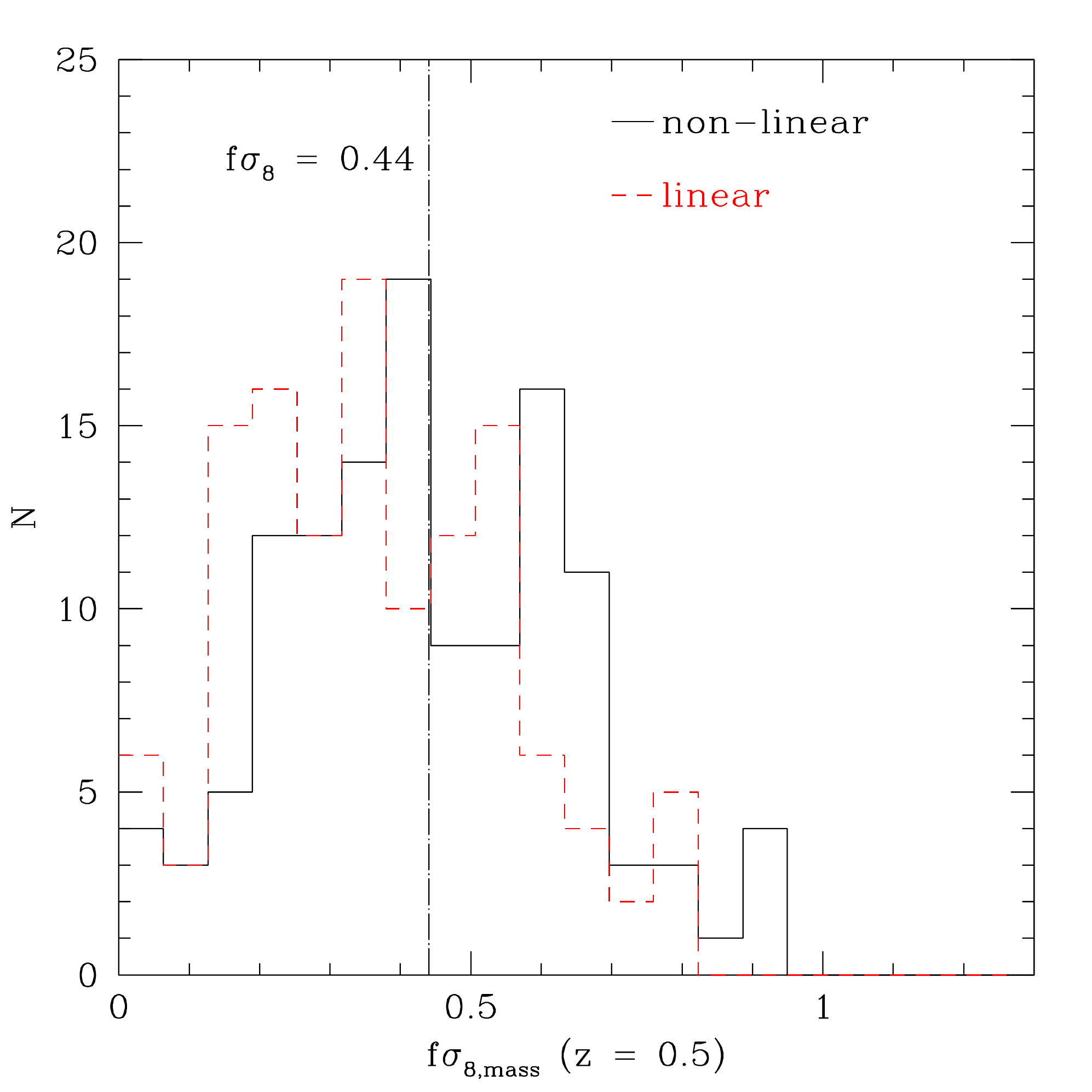}
\caption{Histograms displaying the distribution of $f(z)\sigma_{8}(z)$
  values obtained from MICE mock catalogs with $\bar{z} = 0.5$,
  $\Delta z = 0.05(1+z)$, and $\sigma_z = 0.02(1+z)$, when non-linear
  effects are included in the modelling (solid black) and are not included
  in the modelling (dashed red). \label{fig:histlincom}}
\end{figure}

The MICE simulations obviously include non-linear effects from the gravitational clustering in the distribution of matter, and we need to model these effects in order to
match the recovered angular correlation function. If we do not add non-linear effects to the spatial correlation function (i.e., $\xi(r,z) = D^2(z)\xi_{lin}(r,0)$) when fitting to the data, we obtain a
significantly different distribution of best-fit $f(z)\sigma_{8}(z)$
values. We note that in both cases we are still using the linear model described by Eqs. 2-7 to apply the effects of RSD. The histogram of this distribution for the redshift bin with
$\Delta z = 0.05(1+z)$ and $\sigma_z = 0.02(1+z)$ is displayed
displayed in red in Fig.~\ref{fig:histlincom} (with the original
histogram including non-linear effects displayed in black).  The
distribution is significantly skewed towards low $f(z)\sigma_{8}(z)$
values.  The mean of the distribution decreases to 0.37 (from 0.43) --- a 14 per cent change.

\begin{figure}
\includegraphics[width=84mm]{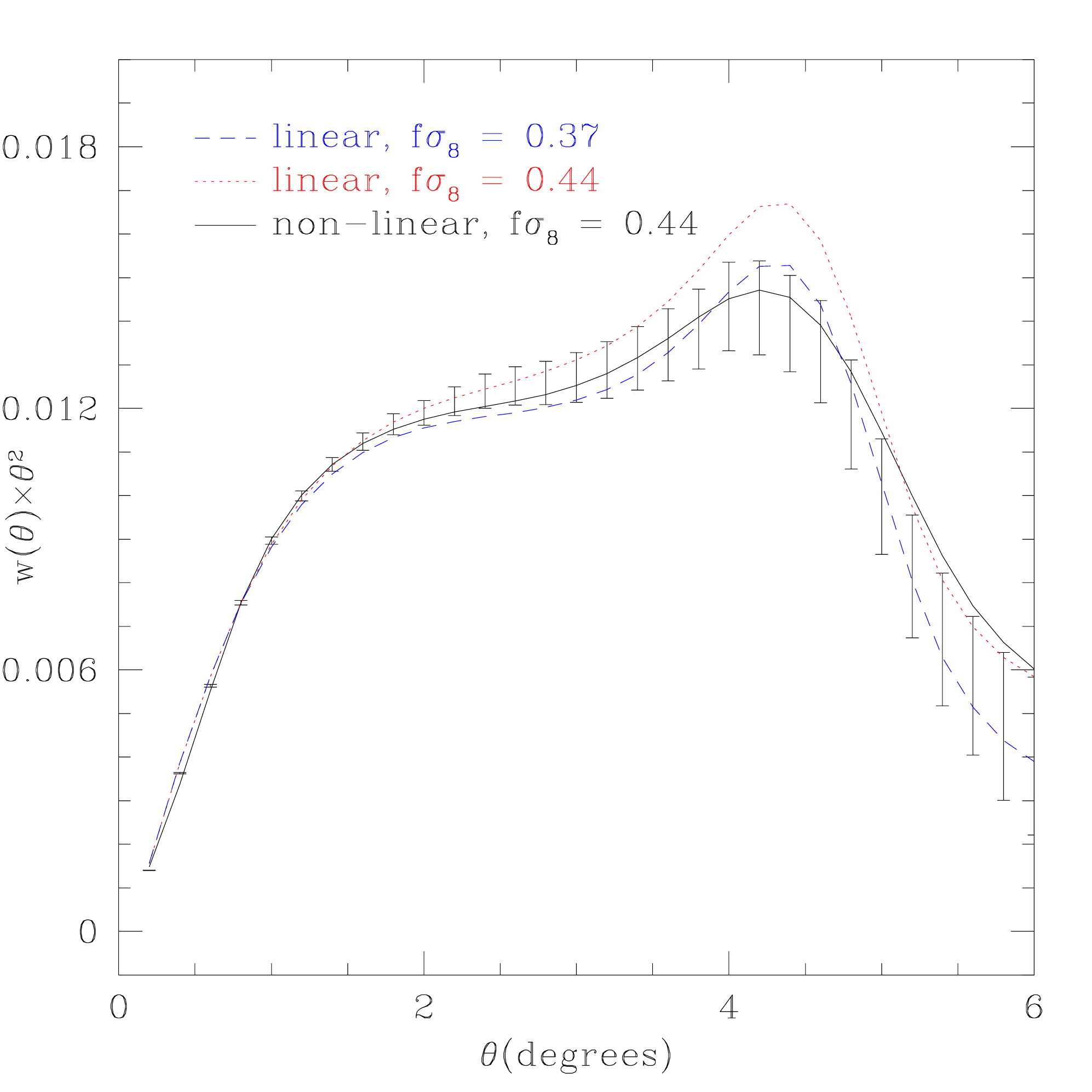}
\caption{The mean angular correlation function from the MICE mocks
  for a redshift bin centred on $z = 0.5$, $\Delta z = 0.05(1+z)$, and
  $\sigma_z = 0.02(1+z)$ is shown by the black error-bars.  The solid black curve
  displays the model $w(\theta)$ including non-linear effects and the
  LCDM prediction of $f(0.5)\sigma_{8}(0.5) = 0.44$.  The dotted red curve
  displays the model when non-linear effects are not included and
  $f(0.5)\sigma_{8}(0.5) = 0.44$ and the dashed blue curve displays the model
  when non-linear effects are not included and $f(0.5)\sigma_{8}(0.5)
  = 0.37$ (that is the best-fit when non-linear effects are not
  included). }
\label{fig:w2lincom}
\end{figure}

Fig.~\ref{fig:w2lincom} shows model $w(\theta)\times\theta^2$
(curves) and the mean $w(\theta)\times\theta^2$ from the MICE mocks
(black error-bars) for a redshift bin centred at $z = 0.5$, $\Delta
z(1+z) = 0.05$, and $\sigma_z = 0.02$.  The black curve displays the
model including non-linear effects and $f(0.5)\sigma_{8}(0.5) = 0.44$
(that is the standard $\Lambda$CDM prediction at $z = 0.5$).  The red
curve displays the model without non-linear effects and
$f(0.5)\sigma_{8}(0.5) = 0.44$, and the blue-curve displays the same
model for $f(0.5)\sigma_{8}(0.5) = 0.37$ (that is the best-fit when
the non-linear effects are not included).  Essentially, the non-linear
effects soften the BAO peak, and when they are not included in the
models, the data is best-fit by a model with a smaller
$f(z)\sigma_{8}(z)$ in order to compensate. We note that the error-bars displayed are not the standard deviations, $\sigma_{M}$, of the mocks but the errors in their mean (i.e., $\frac{1}{\sqrt{N}}\sigma_{M}$ and there are $N=125$ mock realizations). Thus, the disagreement at large scales between the model (black curve) and the mocks is at the $\sim 1\sigma$ level. Given that the $f(z)\sigma_8(z)$ values determined from the mocks agree with the analytic predictions, this disagreement at large scales is too slight to significantly bias our results.   

We find that the best-fit value of $f(0.5)\sigma_{8}(0.5)$ has
shifted by 14 per cent when we do not include non-linear effects. The main non-linear contribution is due to the damping of the BAO due to large-scale velocity flows. This effect thus grows as the Universe evolves (see, e.g., Appendix A of C10).  This implies that we expect the linear theory to yield a less biased result at higher redshift.  This is indeed the case.  We can use our theoretical predictions to find the linear model $w(\theta)$ that is the best-fit to the non-linear model $w(\theta)$.  At $z = 0.7$, we find decreasing the value of $f(0.7)\sigma_{8}(0.7)$ by 8 per cent best matches the non-linear model, and that at $z = 1.0$, decreasing the value of $f(1.0)\sigma_{8}(1.0)$ by 6 per cent yields the best-fit to the non-linear model.  While the magnitude of the bias does decrease with redshift,  it is quite clear that accurate non-linear modelling of $w(\theta)$ is required to obtain accurate values of
$f(z)\sigma_{8}(z)$. 

\section{Measuring Redshift-Space Distortions in the DES}  
  \label{sec:measureRSD}

\begin{figure}
\includegraphics[width=84mm]{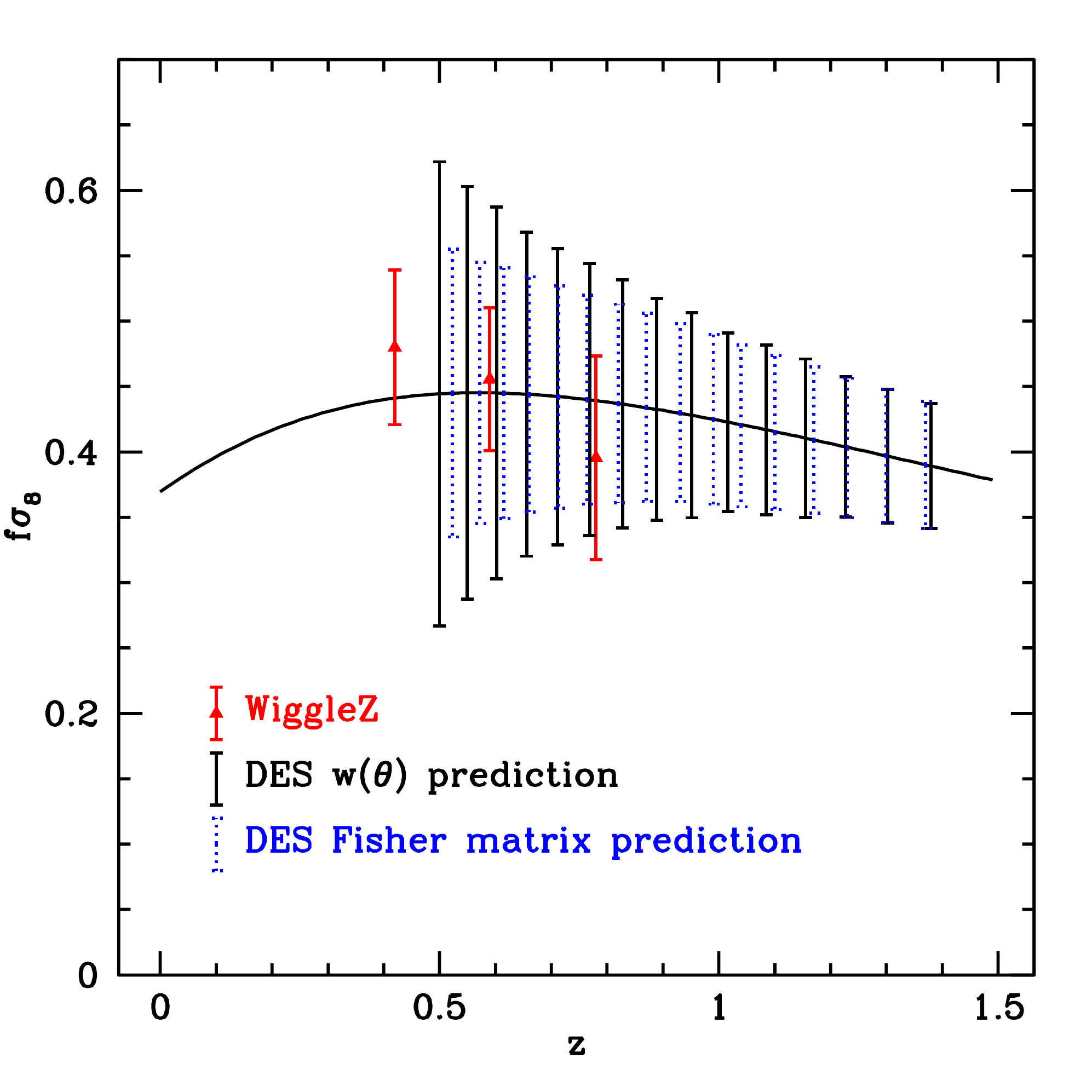}
\caption{The solid line displays the model $f(z)\sigma_{8}(z)$ for our
  default $\Lambda$CDM model. $1\sigma$ errors (black) were calculated
  for the expected measurements made via successive top-hat
  photometric redshift bins for DES galaxies between $0.475<z<1.42$
  and the blue error-bars display Fisher matrix predictions for
  similar redshift bins.  The red points with $1\sigma$ errors are the
  measurements made with the WiggleZ survey
  \citep{blake10}. \label{fig:fpre}}
\end{figure}

Fig.~\ref{fig:TH.5} shows that RSD will significantly affect the
angular clustering measured from DES for top-hat bins. We now quantify
the constraints that could be possible for the DES. The blue
error-bars in Fig.~\ref{fig:fpre} represent the uncertainty that one
obtains from Fisher matrix predictions for the full-anisotropic 3D
power spectrum for the DES. These were calculated using the method
described in \cite{fish}, with a radial damping term for the power
spectrum of the form $e^{-(k\sigma_z\mu)^2}$, where we assume the
photometric redshift error $\sigma_z$ is the dominant cause of radial
damping (and $\mu$ remains the cosine of the angle to the line of
sight).  These predictions are for successive bins of width 0.03(1+z),
the minimum redshift of the first bin being 0.5, and $\sigma_z =
0.03(1+z)$.

We now consider how closely we can match the Fisher matrix errors
using angular clustering measurements. We shall use the angular
correlation function for pair centre binning, $w_{pc}$, which is independent of
RSD, to measure $b(z)\sigma_{8}(z)$. The angular correlation
function for top-hat binning depends on both $f(z)\sigma_{8}(z)$ and
$b(z)\sigma_{8}(z)$, so the combination of the two measures both. Thus, despite
 the relative weakness of the signal-to-noise of $w_{pc}$ (see Fig. \ref{fig:omcl}), the fact that $w_{pc}$
 is independent of $f(z)\sigma_{8}(z)$ allows us to break the $f(z)\sigma_{8}(z)$ -
$b(z)\sigma_{8}(z)$ degeneracy inherent in most clustering measures. (Though, it should be noted
that the Fisher matrix prediction does not include the information about pairs outside of 
the photometric redshift slice.)

We assume that all binned correlation function measurements are drawn
from a multi-variate Gaussian distribution, with covariance matrices
calculated as described in Section~\ref{sec:cov}. Thus, we calculate the standard $\chi^2$ statistic, constructing the full covariance matrix from those of $w_{TH}$ (Eq. \ref{eq:clcov}) and $w_{pc}$ (Eq. \ref{eq:vpc}) and estimating ${\rm Cov}_{TH,pc}$ as described at the end of section \ref{sec:pcerr}.
We are therefore able to find the range of $f(z)\sigma_{8}(z)$ values that take up 68 per
cent of the likelihood space around the Likelihood Maximum, while marginalising over the uncertainty to which the bias can be measured.   

We constrain the range of angles used in these constraints to be such
that $20 h^{-1}$ Mpc$ < r_{eq} < 100 h^{-1}$ Mpc, where $r_{eq} =
2x(\bar{z}) {\rm tan}(\theta/2)$ (with $\theta$ in radians) and $x(z)$
is the co-moving distance to the median redshift, $\bar{z}$, of sample
of galaxies we consider (this has been shown by, e.g., C10 to be an appropriately small scale to which the modelling is still accurate). We use 15 contiguous top-hat
photometric redshift bins, and 15 matched pair-centre bins between
$0.475<z<1.42$, each of width $0.0333(1+z)$, so the first bin has
$0.475<z<0.525$ and the last $1.34<z<1.42$. (The fact that Fig. 1 of C10 suggests the mean photometric redshift uncertainties are expected to $\sim$ constant over the range $0.45 < z < 1.4$ justifies our chosen redshift range.) We conservatively assume
that we can select 7 million galaxies in each redshift bin, with good
photometric redshifts ($\sigma_z \sim 0.03(1+z)$) out of a total of
300 million galaxies: the approximate DES galaxy density distribution
shows that we should expect more than 10 million galaxies in each bin
in total.

The expected constraints from these angular clustering measurements
are shown in Fig.~\ref{fig:fpre}, compared with the Fisher matrix
predictions. At high redshift, the predictions are nearly identical,
but at low redshift the Fisher matrix predicts better constraints than
using the angular clustering method described above. This is
considered further in the next section, where we determine that the
optimal bin at $z \sim 1$ is $\sim 0.03(1+z)$, but at $z \sim 0.5$, a
narrower bin recovers better constraints. We have also plotted recent
measurements made using the WiggleZ survey \citep{blake10} in
Fig.~\ref{fig:fpre}. WiggleZ already does better
than DES at its redshifts, but beyond the redshift limits of WiggleZ ($z \sim 0.8$), DES will provide competitive measurements with better than $\sim$20$\%$ error for slices of width $\Delta
z=0.03(1+z)$. We note that the $f(z)\sigma_{8}(z)$ constraints between
separate redshift bins are highly correlated with each other. We take these
correlations into account when determining how well the DES constrains $f(z)\sigma_{8}(z)$
over its full redshift range in section \ref{sec:comz}.

\subsection{Optimizing the Method}  \label{sec:gal_sample}

The size of the effect of RSD on angular clustering measurements is
strongly dependent on the characteristics of the sample, and in
particular the true radial size of the redshift slice is one of the
most important factors.  This is given by a convolution between the
photometric redshift bounds of the bin and the photometric redshift
distributions. The mean bias of the galaxies is also important, as the
relative amplitude (compared to the real-space clustering) of the
redshift distortion effect is inversely proportional to the bias.

\begin{figure}
  \includegraphics[width=84mm]{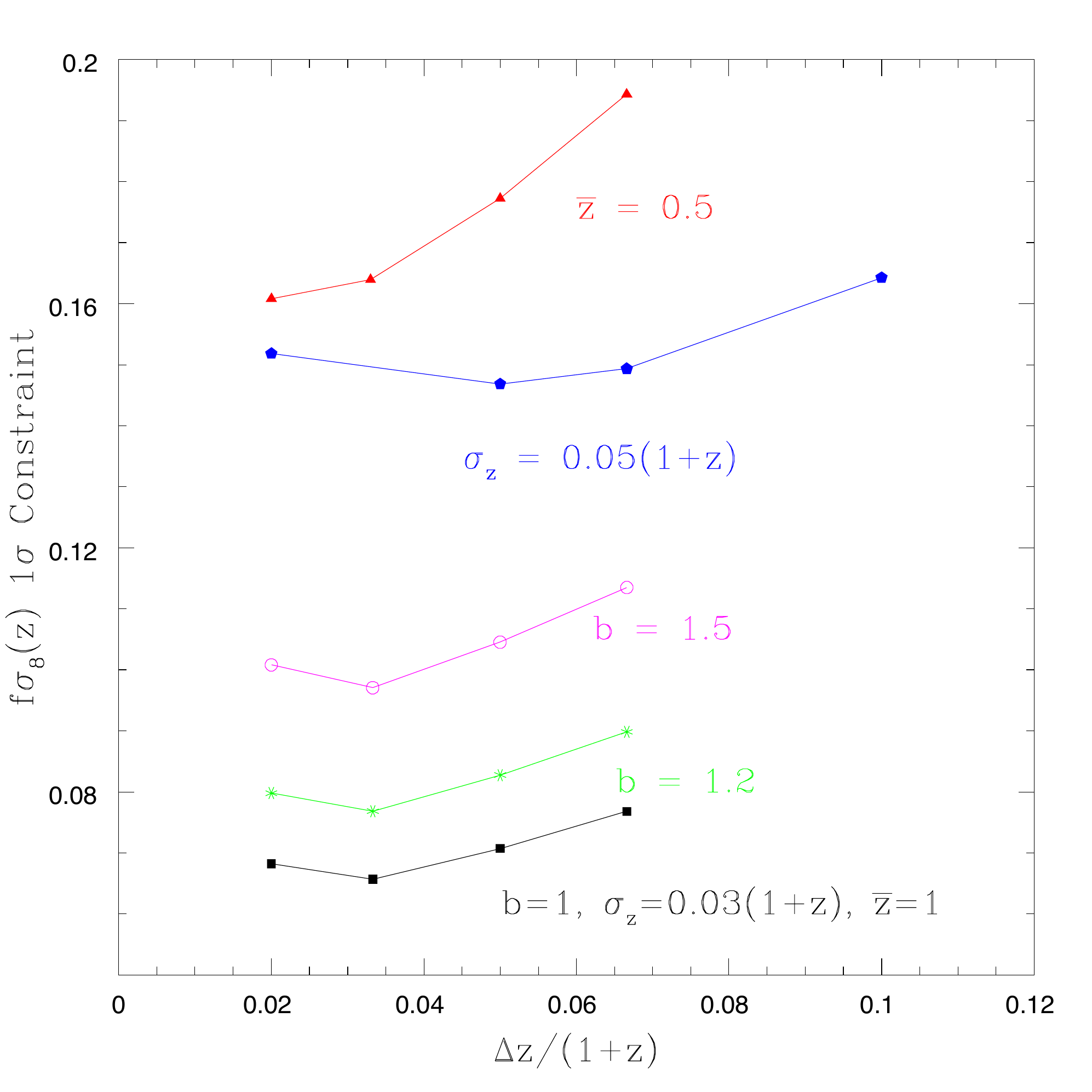}
  \caption{ The black points show the expected error on
    $f(z)\sigma_{8}(z)$ versus the width of the photometric redshift
    bin, for unbiased tracers with average photometric redshift error
    $\sigma_z = 0.03(1+z)$, selected from redshift bins centred on $z
    = 1.065$.  The other lines display the same information, for
    samples changed as labeled. \label{fig:binw}}
\end{figure}

Fig.~\ref{fig:binw} displays the expected error on
$f(z)\sigma_{8}(z)$, for various possible data samples.  The black
line is for unbiased tracers centred at $z=1.065$ and with $\sigma_z =
0.03(1+z)$, which gives a minimum in the expected error for the bin $\Delta
z(1+z) = 0.033(1+z)$.  This represents a balance between the number of
galaxies in each sample against the size of the RSD signal. The blue
line shows the effect of a sample selected with larger photometric
redshift errors, now $0.05(1+z)$.  The expected error on
$f(z)\sigma_{8}(z)$ is more than twice as high, and the minimum is at
a larger bin width.  The true bin width is determined by a convolution
between the photometric redshift bounds of the bin and the photometric
redshift distributions so decreasing the photometric redshift limits
has less of an effect when $\sigma_z$ is increased.

Fig.~\ref{fig:binw} also shows that the effect of increasing $b$ has a
detrimental effect on the RSD measurements: a 50 per cent increase in
the average galaxy bias results in a $\sim$ 50 per cent increase in
the predicted RSD error.  (The error on $\beta = f/b$ remains
constant, implying the error $f(z)\sigma_{8}(z)$ is proportional to
the bias.)  This is an important consideration for the selection of
galaxy samples, particularly at high redshifts where the samples have
larger biases (see, e.g., \citealt{ZZ08}), as do red galaxies that
currently have the lowest photometric redshift errors.

At lower redshifts, Fig.~\ref{fig:binw} shows that RSD measurements
are strongly degraded by the significantly smaller volume available.
At these redshifts, the error decreases significantly with bin size,
down to the most narrow bin we test, $\Delta z(1+z) = 0.02(1+z)$,
though it does appear to asymptote to $\sim0.16$.  This is due to the
fact that the number densities are so high at this redshift that
decreasing the total number of galaxies has a minimal effect on the
error on $w(\theta)$. It is clear from Fig.~\ref{fig:binw} that the
choosing the optimal sample with which to measure RSD depends on a
complicated balance of many interdependent factors: this can only
reliably be achieved after all of the data is in hand.

\subsection{Cosmological Constraints from full Redshift Range}
\label{sec:comz}
We now consider cosmological constraints from the combination of
measurements at different redshifts. We account for the covariance in $f(z)\sigma_8(z)$ between redshift bins by assuming that their Pearson coefficient is equal to the Pearson coefficient that defines the covariance of $w(\theta)$ between different redshift bins. We determine the covariance of $w(\theta)$ between different redshift bins by assuming (see, e.g., \citealt{Thomas10})
\begin{equation}
{\rm Cov}(C_{\ell}^{i},C_{\ell}^{j}) = \frac{1}{(2\ell+1)} (C_{\ell}^{i,j})^2
\end{equation}
and then substituting ${\rm Cov}(C_{\ell}^{i},C_{\ell}^{j})$ for $\sigma^2(C_{\ell}^{i,j})$ in Eq. \ref{eq:clcov}, yielding ${\rm Cov}(w(\theta_1)^{i},w(\theta_2)^{j})$. We find that the Pearson coefficient given by ${\rm Cov}(w(\theta_1)^{i},w(\theta_2)^{j})/\sqrt{{\rm Cov}_{i,i,\theta_1,\theta_2}{\rm Cov}_{j,j,\theta_1,\theta_2}}$ is nearly independent of $\theta_1, \theta_2$. Thus, we set 
\begin{equation}
{\rm Cov}_{f,i,j}/\sqrt{{\rm Cov}_{f,i,i}{\rm Cov}_{f,j,j}} = {\rm Cov}(w^{i},w^{j})/\sqrt{ {\rm Cov}_{i,i} {\rm Cov}_{j,j}}
\label{eq:covf}
\end{equation}
where ${\rm Cov}_{f,i,j}$ is the covariance in $f(z)\sigma_8(z)$ between redshift bins $i$ and $j$. We find that if we make Gaussian realizations of our simulated data matching the covariance matrix for correlation functions expected for two adjacent redshift bins, the covariance between measurements of $f(z)\sigma_8(z)$ recovered is within 10\% of what we estimate using Eq. \ref{eq:covf} (and is not systematically offset). Thus, we use Eq. \ref{eq:covf} to construct our full covariance matrix for $f(z)\sigma_8(z)$, given its relative ease in computation and that it should be sufficiently accurate for our tests.

\subsubsection{Constant offset for $f(z)\sigma_{8}(z)$}

If we consider deviations around our fiducial model with a
redshift-independent offset in $f(z)\sigma_{8}(z)$, then the 68\%
confidence interval for $\Delta f(z)\sigma_{8}(z)$ is given by
\begin{equation}
\Delta \left(f(z)\sigma_{8}(z)\right) = 1/\sqrt{\sum_{i,j}{\rm Cov}^{-1}_{f,i,j}}.
\end{equation} 
\begin{figure}
  \includegraphics[width=84mm]{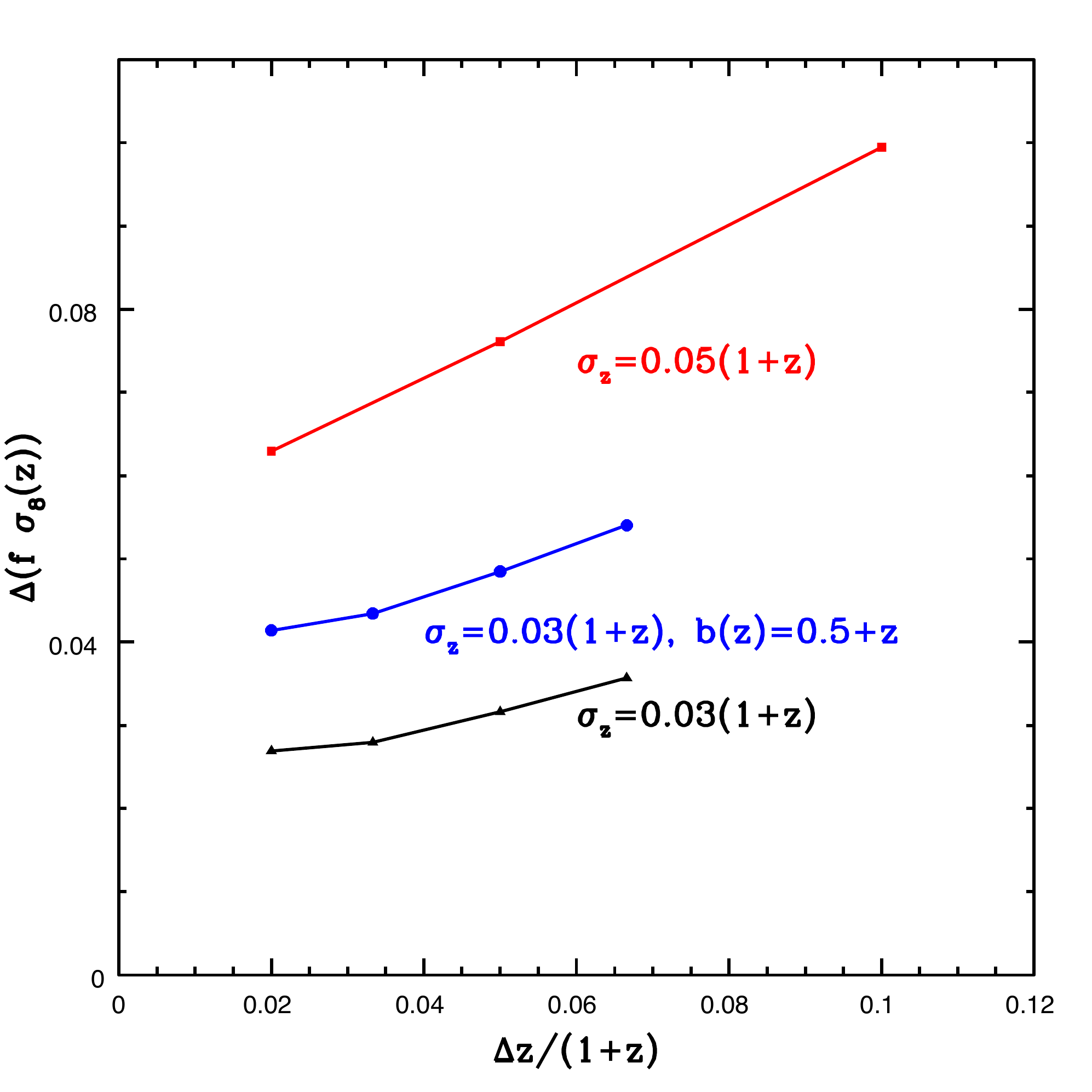}
  \caption{ 68\% confidence limits for a constant offset in
    $f(z)\sigma_{8}(z)$ from our fiducial model, when we combine the
    constraints produced by a series of $w_i(\theta)$ measured in
    successive redshift bins with median redshifts between 0.5 and
    1.4, versus the width of the photometric redshift bins.  The black
    symbols show the prediction for unbiased tracers with average
    photometric redshift error $\sigma_z = 0.03(1+z)$.  The red points
    display the same information for $\sigma_z = 0.05(1+z)$ and the
    blue points assume that the bias of the selectable galaxies
    evolves, such that $b(z)=0.5+z$ and the average photometric
    redshift error of these galaxies is $\sigma_z =
    0.03(1+z)$.\label{fig:fsigcons}}
\end{figure}
We determine $\Delta \left(f(z)\sigma_{8}(z)\right)$ for unbiased
tracers occupying a variety of bin widths (all with the median
redshift of the first bin at 0.5 and the median redshift of the final
bin less than 1.4; the number of galaxies in each bin is assumed to be
7$\times 10^6\times(\Delta z(1+z)/0.05)$) and plot the results in
Fig.~\ref{fig:fsigcons}. We find that, for $\sigma_z = 0.03(1+z)$,
$\Delta \left(f(z)\sigma_{8}(z)\right)$ asymptotes towards a value of
0.025.  For $\sigma_z = 0.05(1+z)$ the decrease in the error as a
function of bin width is much stronger, but at the minimum bin width
we probe ($0.02(1+z)$) we find $\Delta \left(f(z)\sigma_{8}(z)\right)
= 0.06$ --- more than twice as large as for $\sigma_z = 0.03(1+z)$.

We also investigate a scenario with more realistic evolution in the
bias of the galaxies selected, assuming $b(z) = 0.5+z$.  This allows
the bias to be unity at $z=0.5$ and gives a similar value to the bias
of galaxies in the DEEP2 field (see, e.g., \citealt{ZZ08}) at $z = 1$.
For this bias model, we find $\Delta \left(f(z)\sigma_{8}(z)\right)$
asymptotes towards 0.04.  Overall, this suggests that DES alone should
be able to rule out models that cause the growth history to differ by
more than 10 per cent from the $\Lambda$CDM prediction.  It further
suggests that, though the covariance between bins grows large (the
reduced covariance of off-diagonal terms is as large as 0.9), extra
information can be gained by making the bins as narrow as possible.

\subsubsection{Deviations in $\gamma$}

\begin{figure}
  \includegraphics[width=84mm]{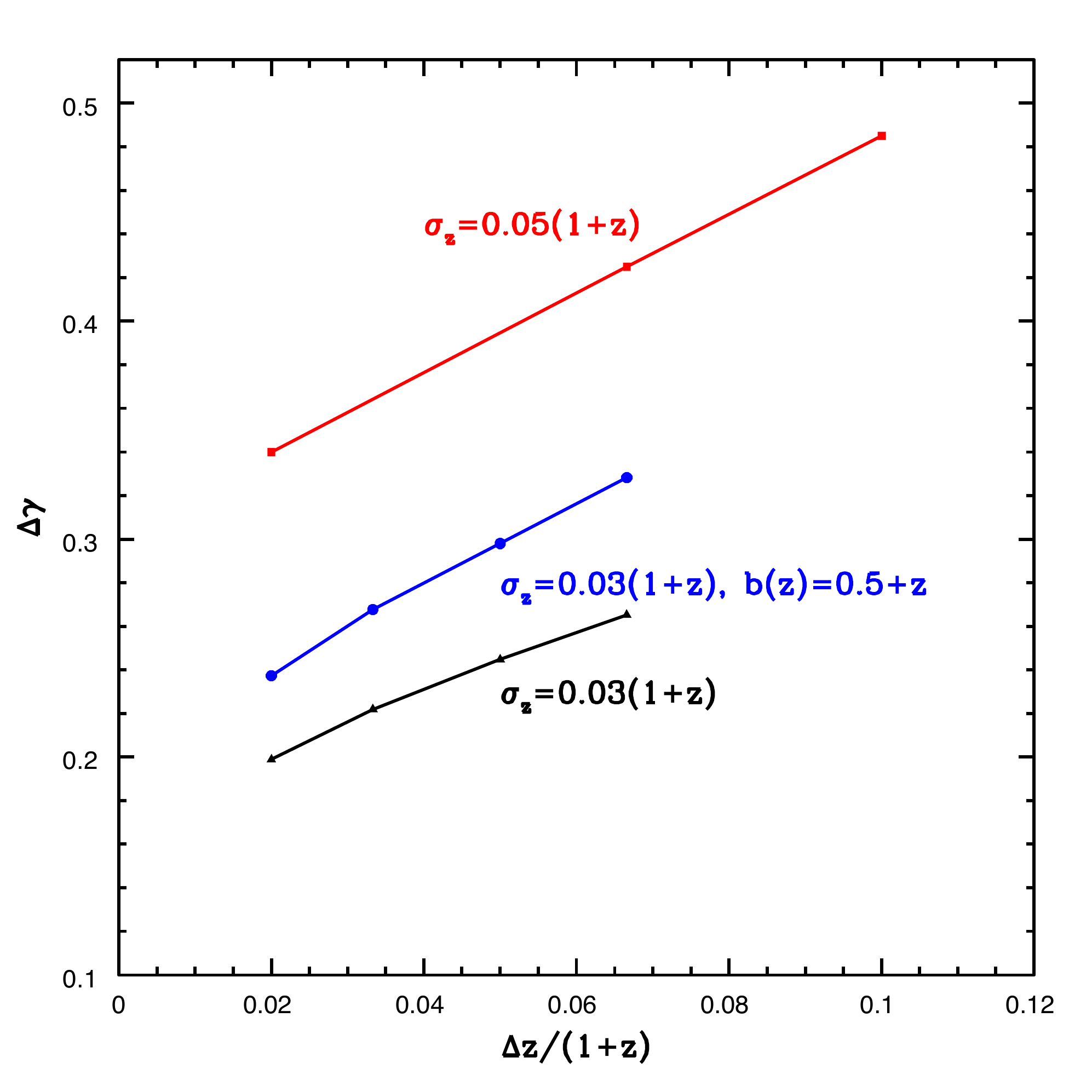}
  \caption{As Fig.~\ref{fig:fsigcons}, but now showing the expected
    constraints on the value of $\gamma$, given the model $f(z) =
    \Omega_m(z)^{\gamma}$. \label{fig:gamcons}}
\end{figure}

One can also determine how well DES will be able to constrain models
of the type $f(z) = \Omega_m(z)^{\gamma}$ (see, e.g., \citealt{linder05}).  We note that this also
implies a change in the behaviour of $\sigma_{8}(z)$, since
$\sigma_{8}(z) = \sigma_{8}(0)G(z)$ and $f(z) = {\rm d log}(G)/{\rm d
  log}(a)$.  We fix the value of $\sigma_{8}(1088)$ that yields
$\sigma_{8}(0) = 0.8$ in the $\Lambda$CDM cosmology and determine the
expected behaviour of $f(z)\sigma_{8}(z)$.  We then test this against
the $f(z)\sigma_8(z)$ covariance matrix we obtain for a series of
$w_i(\theta)$ measurements to predict the ability of DES to constrain
$\gamma$.  The results we find are presented in
Fig.~\ref{fig:gamcons}.  We find there is a steady decrease in the
expected error on $\gamma$ as the bin width deceases, and we find that
DES alone should be able to place 1$\sigma$ constraints on the value
of $\gamma$ to less than 0.25, even with realistic evolution in the
bias of the observed galaxies.

\section{Robustness to Underlying Cosmology and Photometric Redshift Systematics}
\label{sec:pc_bin}

The RSD constraints presented in the previous section assumed perfect
knowledge of cosmological geometry and photometric redshift
errors. Both changing the assumed cosmology and introducing systematic
errors can cause scale dependent effects that could mimic the effects
of redshift-space distortions when measuring $w(\theta)$ using a
top-hat bin.  

Measurements of the angular correlation function only depend on
angles and redshifts, and are therefore independent of cosmological
model: a model does not have to be specified in order to make these
measurements. The cosmological dependence only arises through the
model to be fitted to the data. We test the systematic effect of assuming an incorrect
background model on $f(z)\sigma_8(z)$ measurements by using the $1 < z_{phot} < 1.13$ redshift bin. In effect, we consider fitting to a fixed $\Lambda$CDM model with free structure growth, and determine the degree to which the inferred structure growth changes when the assumed $\Lambda$CDM parameters are incorrect. 

We calculate the value of $f(z)\sigma_{8}(z)$ one measures if one assumes the geometrical cosmology given by $\Omega_m = 0.25, f_b = \Omega_b/\Omega_m = 0.18, h=0.7$, but the Universe has a true
cosmology that is different.  For $(f_b)_{true} = 0.2$, the best-fit value of
$f(z)\sigma_{8}(z)$ increases to 0.52$\pm$0.07 (from the true value of 0.42$\pm0.07$)
while it decreases to $f(z)\sigma_{8}(z) =0.31^{+0.09}_{-0.08}$ for $(f_b)_{true} = 0.16$.
Changing $\Omega_m$ produces an even more significant effect, as
the $f(z)\sigma_{8}(z)$ decreases to $0.23^{+0.09}_{-0.10}$ when $(\Omega_m)_{true} = 0.3$.
While these changes to the cosmology are within the 95\% confidence limits determined by WMAP \citep{komatsu09}, by the time that DES finishes, the Planck
Experiment will have set tighter constraints on these parameters, making this effect less important. However, given current
constraints, a joint fit to all of the parameters is clearly required.

There are other effects that can cause scale dependent changes to the observed $w(\theta)$. Any scale dependent bias will clearly be a problem, and it has recently been shown that large halos display scale dependent biases (see, e.g., \citealt{smith08,man09,des10}). However, as shown in Fig. 8, the best RSD constraints will be obtained from objects that reside in the least biased halos, for which we expect the least scale-dependent bias. Any non-zero $f_{NL}$ will also cause a scale dependent bias, though for $|f_{NL}| \sim 100$, this is a much smaller effect than RSD at scales smaller than the BAO scale. Additionally, we may worry about observational systematics, such as Galactic extinction or stellar contamination. \cite{R11} has shown that a thorough accounting of and correction for such systematics and their effects is possible, but depending on the particular survey, this could translate to additional systematic errors in estimation of $f(z)\sigma_8(z)$. We note that the systematic effects are largest at scales greater than the BAO scale, and are therefore less likely to strongly affect measurements of $f(z)\sigma_8(z)$.

\subsection{Incorrect Redshift Distributions}
\label{sec:bdndz}
One might worry that the determination of $f(z)\sigma_{8}(z)$ (or any
other cosmological parameter) will be particularly sensitive to
mis-estimation of the true redshift distribution of galaxies within a
particular redshift bin.  In order to investigate this, we alter the
$\sigma_z$ and also simulate the effect of a redshift dependent bias
in our modelling. Previously, we assumed $\sigma_z = 0.03(1 + z)$ for
the DES galaxies, and we now contrast against models calculated
increasing or decreasing this assumed error by 10$\%$. This is about
the same as the average uncertainty in $\sigma_z$ stated in the DES
requirements document, which states that the uncertainty in the
dispersion in any photometric redshift bin of width $0.1(1+z)$ should
be less than 0.003.

\begin{figure}
\includegraphics[width=84mm]{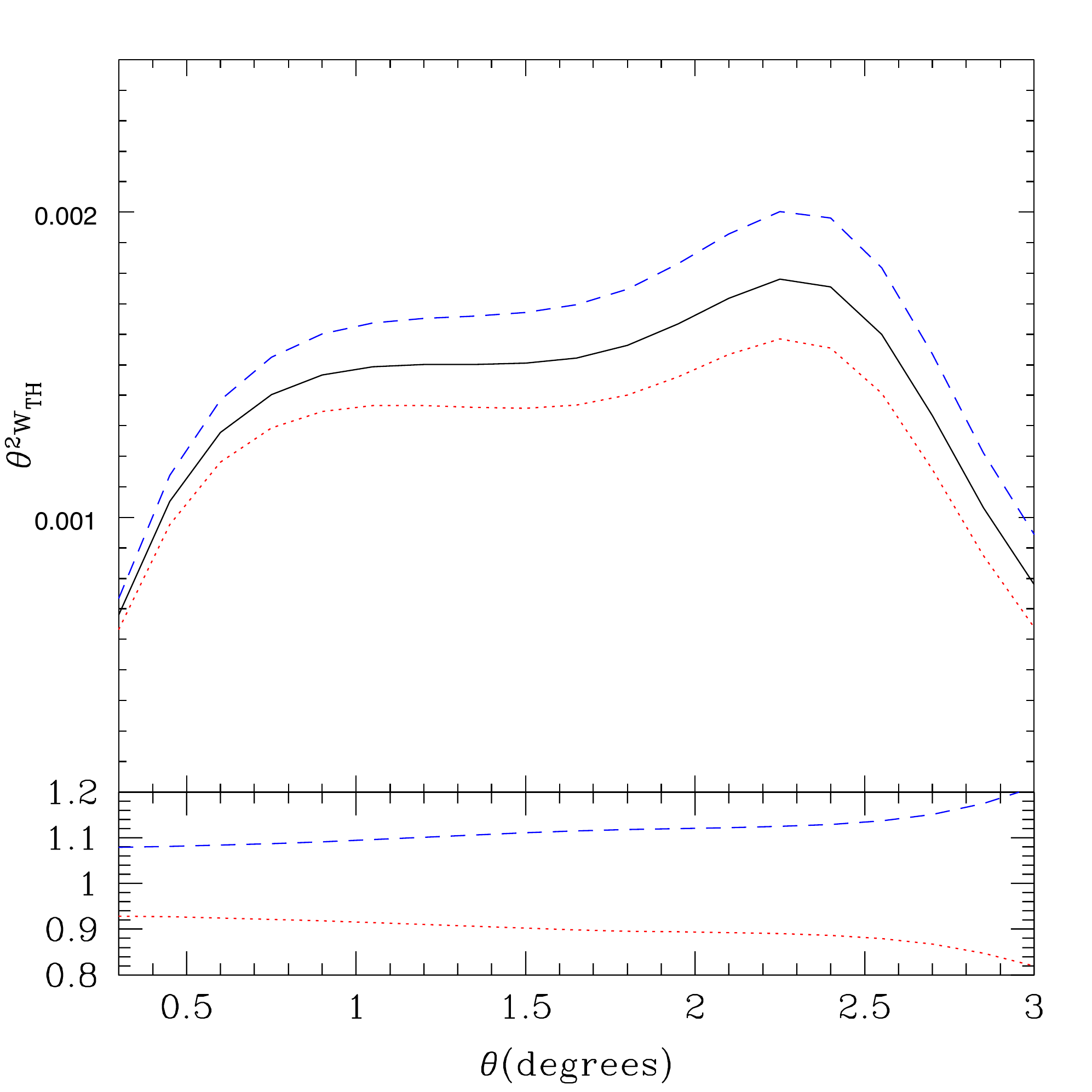}
\caption{The top panel displays the model $w(\theta)$, multiplied by
  $\theta^2$ from the top-hat bin defined by $1.0 < z < 1.13$ with
  $\sigma_z = 0.03(1+z)$ (black line), $\sigma_z = 0.033(1+z)$ (red
  dotted line) and $\sigma_z = 0.027(1+z)$ (blue dashed line).  The bottom panel
  displays the ratio of the models with $\sigma_z = 0.033(1+z)$ and
  $\sigma_z = 0.027(1+z)$ to the model with $\sigma_z =
  0.03(1+z)$. }
\label{fig:zerrTH}
\end{figure}

\begin{figure}
\includegraphics[width=84mm]{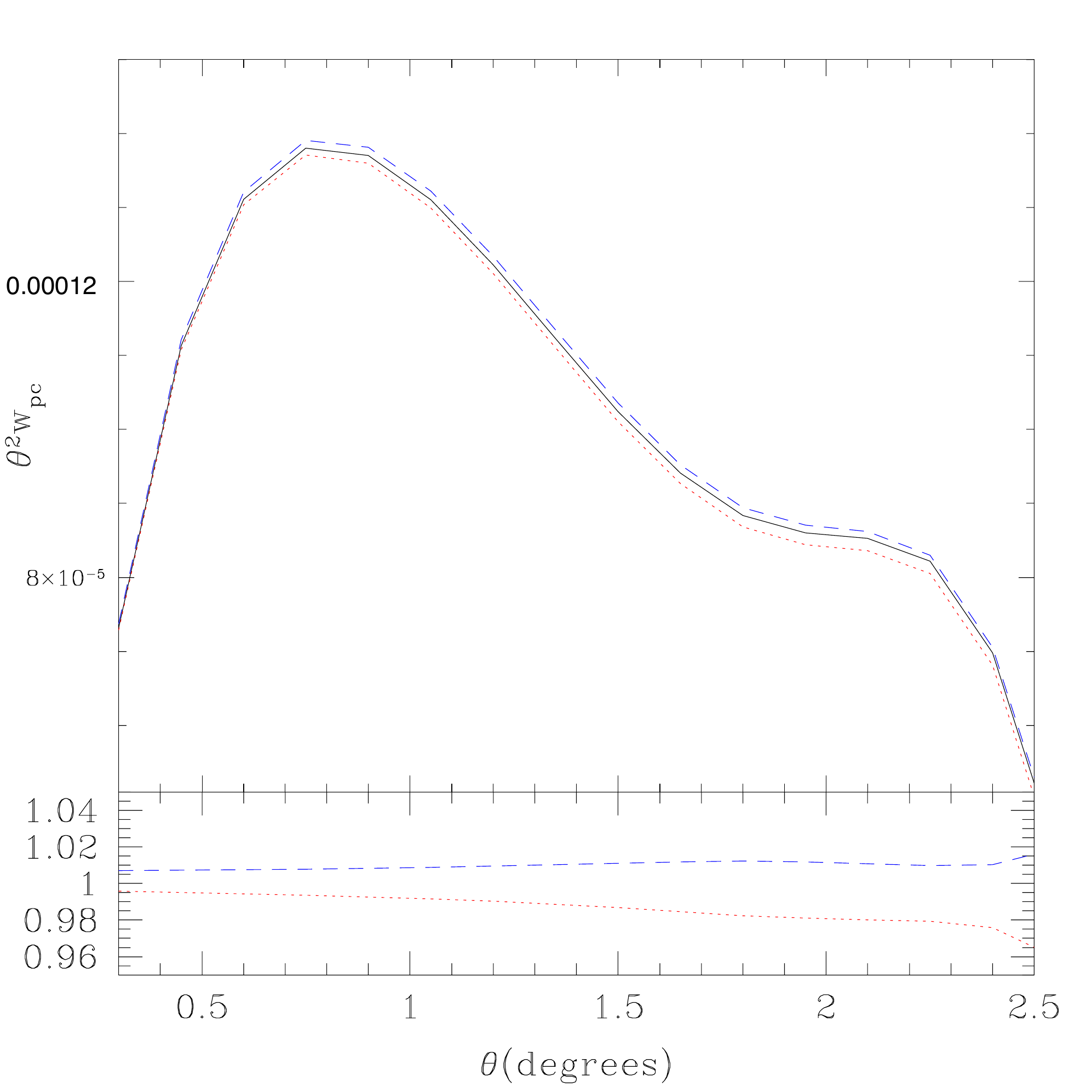}
\caption{The model pair-centre measurements for the three situations
  described in Fig.~\ref{fig:zerrTH}. }
\label{fig:zerrpc}
\end{figure}

\begin{figure}
\includegraphics[width=84mm]{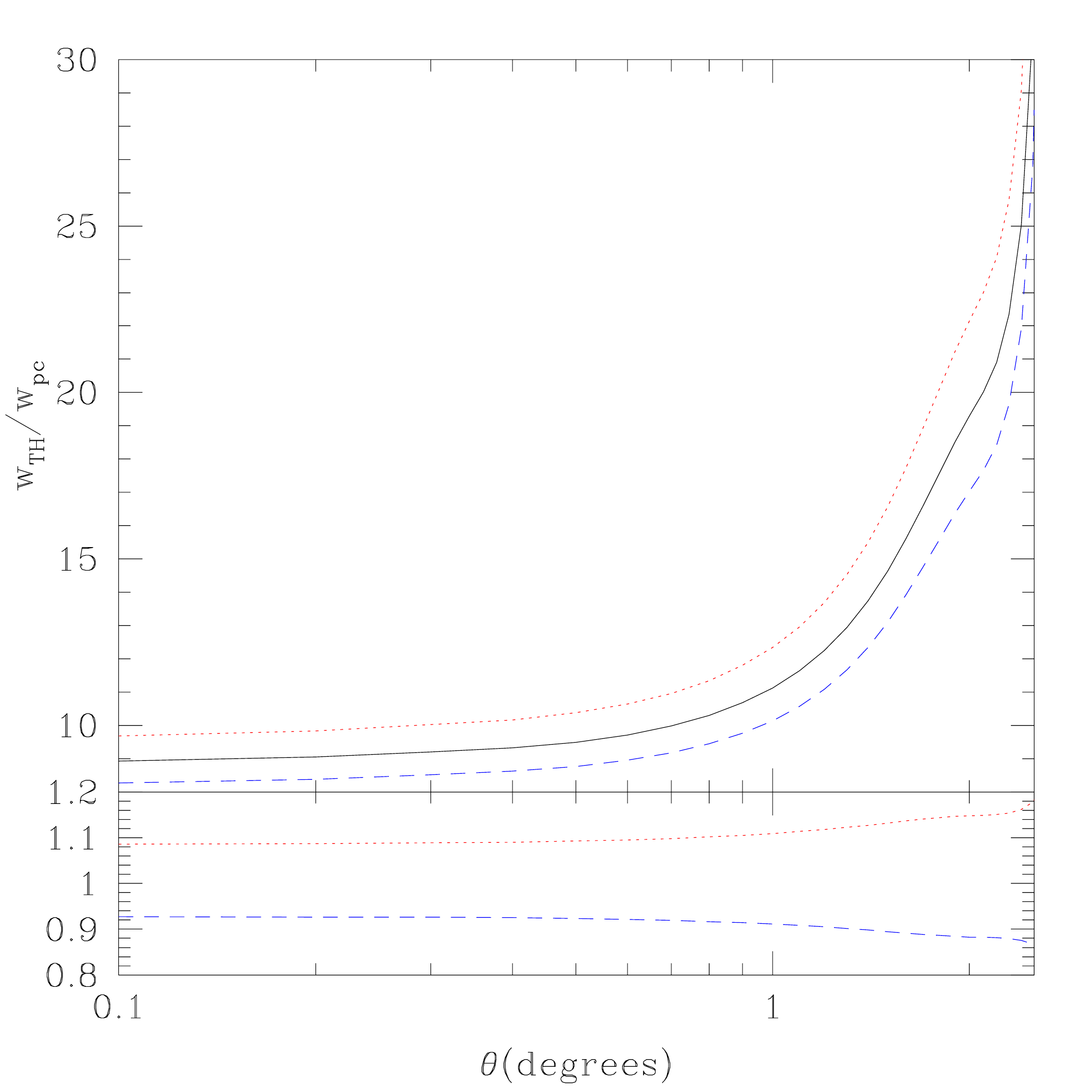}
\caption{Same as Fig.~\ref{fig:zerrTH}, but the ratio between the
  model top-hat and model pair-centre measurements are plotted
  instead, and the $\theta$ axis is scaled logarithmically. }
\label{fig:zerr}
\end{figure}

The effect on the top-hat measurement of changing $\sigma_z$ is shown
if Fig.~\ref{fig:zerrTH}.  Its amplitude changes by $\sim$ 10 per
cent, but its shape remains very similar. The bottom panel of Fig. \ref{fig:zerrTH}, shows that there is a slight scale-dependency in the change in amplitude, similar to what C10 find (as displayed in their Fig. 6). Figs.~\ref{fig:zerrpc} and~\ref{fig:zerr} display the same information as Fig. \ref{fig:zerrTH} for the  pair-centre measurement and the ratio of the
top-hat measurement to the pair centre
measurement, respectively.  The pair-centre measurement displays
almost no dependence on $\sigma_z$.  The pair-centre measurements
change by less than 2$\%$, and the change in the top-hat bin is closer
to 10$\%$.  Further, at small scales, where the redshift distortion
effect is relatively small, the ratio between the different
pair-centre measurements differs by $<$ 0.5 $\%$.  This implies that the
pair-centre measurement would continue to allow for an accurate
measure of $b(z)\sigma_{8}(z)$.  However, this value of $b(z)\sigma_{8}(z)$ would
cause the model top-hat $w(\theta)$ to differ by $\sim$ 10 per cent
from the measured $w_{TH}(\theta)$ on scales $< 20h^{-1}$Mpc,
regardless of the value of $f(z)\sigma_{8}(z)$ (since RSD do not have
a significant effect on the amplitude at these scales).

The mismatch between the two measurements on small scales allows one
to discover and fix errors in the estimation of the true redshift distribution, in a way that is relatively
independent of RSD.  Consider the ratio of the top-hat to pair-centre
$w(\theta)$: For the ratio, the values of $b(z)\sigma_{8}(z)$ cancel each
other (and one would expect any non-linear bias to cancel as well).  This implies that even at small scales, where proper modelling of $w(\theta)$ of galaxies requires knowledge of the halo occupation distribution of the galaxies, the ratio would remain constant.  (Note, at large scales, it is the linear effect of RSD that causes the ratio to be non-constant.)  At small-scales, the top-hat and pair-centre measurements should probe the same real-space clustering, i.e., where $w(\theta)$ behaves as a power-law, we would expect the pair-centre and top hat measurements to have the same slope.  
Thus, we expect the ratio to be independent of any non-linear effects.  This implies that, at small scales, any offset between the measured ratio and the theoretical ratio will be due almost entirely to incorrect
modelling of the true redshift distributions.  

At scales where $w(\theta)$ behaves as a power-law, the amplitude of $w(\theta)$ is
proportional to
\begin{equation}
W(\sigma_z) = \int n^2(z,\sigma_z) H(z)dz.
\label{eq:W}
\end{equation}
Thus, one can determine how $\sigma_z$ needs to be changed in order
for $W(\sigma_{z,a})/W(\sigma_{z,m})$ (where $\sigma_{z,a}$ is the
adopted mean photometric redshift error and $\sigma_{z,m}$ is the mean
from the actual data) to match the offset found between the the
measured and predicted ratios between the top-hat and pair centre
measurements.  When the errors are 10 per cent too small, the offset
is 1.083, and we find that changing $\sigma_z$ from 0.0027 to 0.0302
produces a matching $W(\sigma_{z,a})/W(\sigma_{z,m})$.  When the
errors are 10 percent too large, the offset is 0.922, and changing
$\sigma_z$ from 0.033 to 0.0298 produces a matching
$W(\sigma_{z,a})/W(\sigma_{z,m})$.  In both cases, a slight
over-correction was required, but the correction would reduce the
systematic error on any $f(z)\sigma_{8}(z)$ estimate to less than 1
percent.

In practice, one will have a individual photometric redshift errors
(or PDFs) for each galaxy.  In this case, the correction would be to
find the (likely constant) factor by which each error distribution
needs to widened or narrowed in order to achieve the desired ratio in
the model.  This correction is likely to be important for any analysis
one wishes to conduct on the clustering of photometrically selected
galaxies, and it is appealing because does not require any data
external to the survey (unlike other proposed methods, which rely on
smaller spectroscopic surveys, see, e.g. \citealt{Newman08}).

\begin{figure}
\includegraphics[width=84mm]{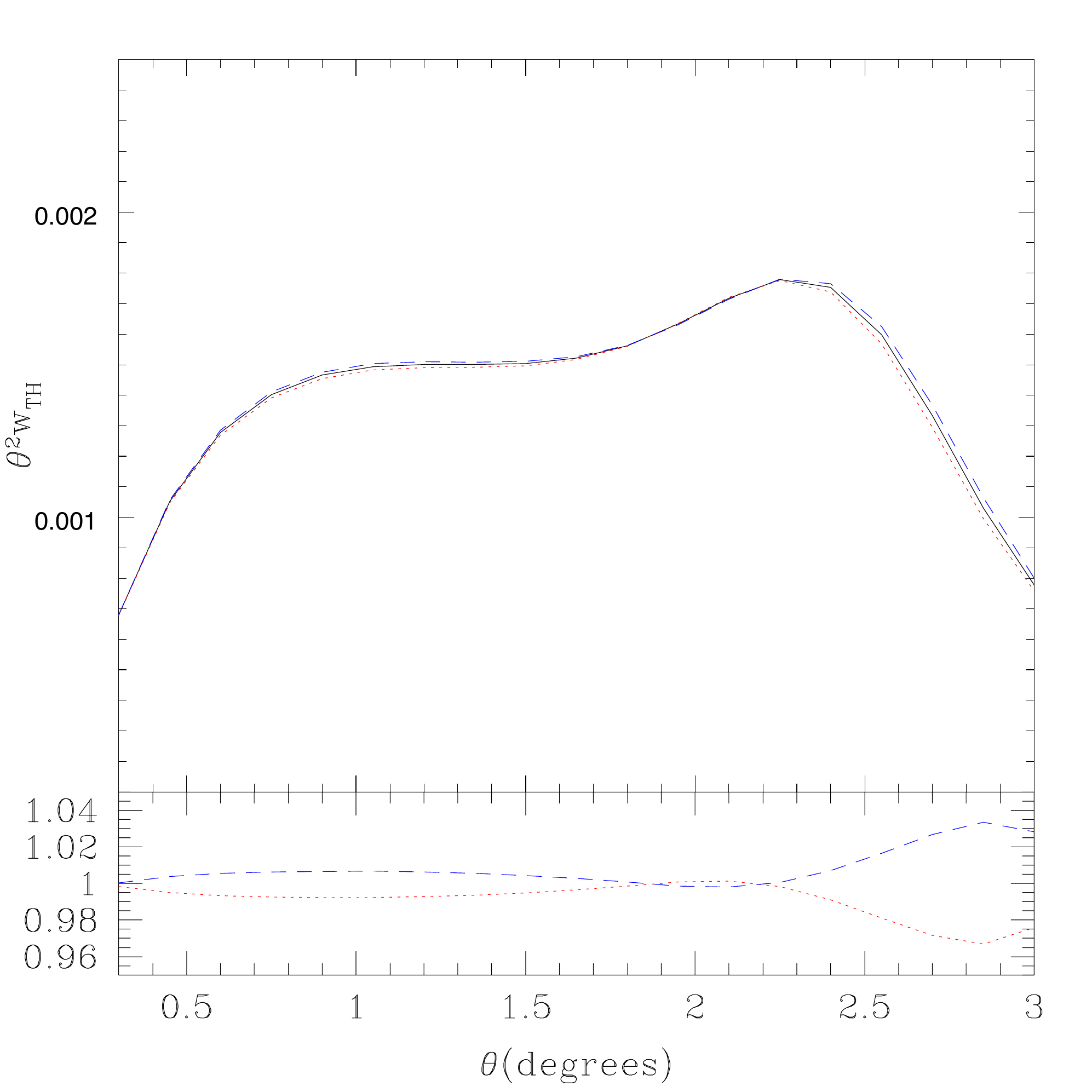}
\caption{ The top panel model $w(\theta)$ for a top-hat bin defined by
  $1.0 < z < 1.13$.  The black line assumes no photometric redshift bias,
  while the red dotted line assumes a bias 0.01, and the blue dashed line assumes a
  bias of -0.01, all for $\sigma_z = 0.03(1+z)$.  The bottom panel
  displays the ratio of the models with bias to the model with an
  unbiased distribution. }
\label{fig:zbias}
\end{figure}

We also investigate the effect of a bias on the photometric redshifts
(i.e., the mean photometric redshift in a bin may be higher or lower
than the mean of the true redshift distribution).  We apply a positive
or negative photometric redshift bias of 0.01.  This level of bias is
close to 10 times higher than the 0.001$(1+z)$ stated in the DES
requirements document.  The effect of these biases is seen in
Fig.~\ref{fig:zbias}.  The model top-hat $w(\theta)$ with no bias is
shown by the black line, while red line is for $z_b = 0.01$, and blue
$z_b = -0.01$. The value of the amplitudes changes only slightly ---
by $\sim$1 per cent.  This is because the main effect is a change in
the median redshift and thus the effective value of the linear growth
rate.  The ratio of $D(1.055)/D(1.075)$ is just 1.01, thus the results
we find match the expectation.  These results suggest that the
requirements of DES will mean that photometric redshift bias is not an
issue for the purposes of measuring $f(z)\sigma_{8}(z)$.

\section{Conclusions}  \label{sec:conc}

We have described the effects of RSD on angular clustering
measurements and outlined an effective method for measuring
$f(z)\sigma_{8}(z)$.  In order to guide these efforts, we have extended
the techniques outlined in C10 to model the amplitude and covariance of $w(\theta)$ measured using galaxies selected from a
DES-like survey, allowing us to determine the optimal binning scheme
to extract the most information about RSD.  In particular:

\noindent$\bullet$ We have shown that RSD have a measurable effect on
the angular correlation function $w(\theta)$ measured using
photometric redshifts. At the very least, any accurate model must
include their effects, or have shown that they are negligible.

\noindent$\bullet$ We describe how using measurements $w(\theta)$ in
pair-centre bins allows $b(z)\sigma_{8}(z)$ to be constrained
independently of RSD, thereby minimizing the degeneracy between
$b(z)\sigma_{8}(z)$ and $f(z)\sigma_{8}(z)$.

\noindent$\bullet$ We have shown that $f(z)\sigma_{8}(z)$ can be
measured to $\sim 17\times b$ per cent accuracy at $z = 1$ for a
DES-like survey, assuming $\sigma_z = 0.03(1+z)$.

\noindent$\bullet$ We confirm that our analytic predictions of the
ability to constrain $f(z)\sigma_{8}(z)$ work well as, when we measure
$f(z)\sigma_{8}(z)$ from mock catalogs generated by the MICE
simulation, we obtain distributions that agree with our analytic
predictions.

\noindent$\bullet$ We show that omitting non-linear gravitational effects around the BAO scale in the
modelling of $w(\theta)$ can cause the expected $f(z)\sigma_{8}(z)$
value one measures to be more than 15 per cent smaller than its true
value.

\noindent$\bullet$ The accuracy with which $f(z)\sigma_{8}(z)$ can be
measured depends strongly on the error on the photometric redshifts.
At redshift 1, the expected error on $f(z)\sigma_{8}(z)$ is $\sim$
twice as large for $\sigma_z = 0.05(1+z)$ compared to $\sigma_z =
0.03(1+z)$.  Given that the expected error depends strongly on the
bias of the galaxies and the photometric redshift limits applied to
the bin, we expect the optimal bin (or series of bins) can only be
determined after the data is in hand.
 
\noindent$\bullet$ We have shown that, after adopting a reasonable
expectation of $b(z)$, a series of $w(\theta)$ measurements made
for galaxies drawn from narrow ($\Delta z(1+z) < 0.02$) redshift
slices over the range $0.5 < z < 1.4$ should allow a DES-like survey
to detect 10 per cent deviations from the $\Lambda$CDM prediction for
cosmological growth of structure.  In terms of the model $f(z) =
\Omega_m(z)^{\gamma}$, we expect a DES-like survey to be able to
determine $\gamma = 0.557^{+0.25}_{-0.22}$.

\noindent$\bullet$ We further show how the ratio can be used to
correct any inaccuracy in the estimation of $\sigma_z$ for a
particular redshift distribution. This provides an attractive method
for testing the photometric redshifts, since it does not require any
external data.

The constraints that we predict will only improve as the redshift
limits and sky area increase with future surveys.  Given purely
photometric data, it is clearly still worth investigating and
measuring the effects of RSD.

\section*{Acknowledgements}
We thank an anonymous referee for comments and suggestions that have improved the quality of this work.

We acknowledge the use of mock catalogues built upon an N-body simulation provided by 
the MICE collaboration (publicly available at http://www.ice.cat/mice).  The MICE simulations have been developed at the MareNostrum supercomputer (BSC-CNS) thanks to grants AECT-2006-2-0011 through AECT-2010-1-0007. Data products have been stored at the Port d'Informaci— Cient'fica (PIC).

AJR and WJP thank the UK Science and Technology Facilities Council for financial support.  WJP is also grateful for support from the Leverhulme trust and the European Research Council.

MC \& EG acknowledge support from the Spanish Science Ministry:
AYA2009-13936, Juan de la Cierva and  Consolider-Ingenio CSD2007-00060,
from Generalitat de Catalunya: project 2009SGR1398 and from the European 
CommissionÕs Marie Curie Initial Training Network CosmoComp (PITN-GA-2009-238356).

\end{document}